\documentclass[a4paper,11pt]{article}
\pdfoutput=1
\usepackage[utf8]{inputenc}
\usepackage{mathtools,amsmath,amssymb}
\usepackage{graphicx}
\usepackage{lmodern}
\usepackage[T1]{fontenc}
\usepackage{microtype}
\usepackage{hyperref}
\usepackage{xspace}
\usepackage{bibentry}

\usepackage[bulletsep]{collref}
\numberwithin{equation}{section}
\usepackage[usenames,dvipsnames,svgnames,table]{xcolor}
\usepackage{tikz}

\usetikzlibrary{shapes}
\usetikzlibrary{decorations.markings}
\usetikzlibrary{decorations.pathmorphing}
\usetikzlibrary{arrows.meta}

     \newcommand{\dcircg}[2]{\draw[color=gray,fill=white] (#1,#2) circle (0.1);}
     \newcommand{\dcrossg}[2]{
        \draw[thick,color=gray] (#1-0.05,#2-0.15) -- (#1+0.05,#2+0.15);
        \draw[thick,color=gray] (#1+0.12,#2-0.07) -- (#1-0.12,#2+0.07);
    }
     \newcommand{\dcirc}[2]{\draw[fill=white] (#1,#2) circle (0.1);}
     \newcommand{\dcross}[2]{
        \draw[thick] (#1-0.1,#2-0.1) -- (#1+0.1,#2+0.1);
        \draw[thick] (#1+0.1,#2-0.1) -- (#1-0.1,#2+0.1);
    }

\usepackage[left=2.6cm,right=2.6cm,top=2.6cm,bottom=3cm]{geometry}

\newcommand{\osc}[1]{\mathbf{#1}}
\newcommand{\oscgreek}[1]{\boldsymbol{#1}}
\newcommand{\dagg}[1]{\bar{#1}}
\newcommand{\oa}{\osc{a}}
\newcommand{\ob}{\osc{b}}
\newcommand{\oc}{\osc{c}}
\newcommand{\od}{\osc{d}}
\newcommand{\oab}{\dagg{\osc{a}}}
\newcommand{\obb}{\dagg{\osc{b}}}
\newcommand{\ocb}{\dagg{\osc{c}}}
\newcommand{\odb}{\dagg{\osc{d}}}
\newcommand{\ox}{\oscgreek{\xi}}
\newcommand{\oxb}{\dagg{\oscgreek{\xi}}}
\newcommand{\och}{\oscgreek{\chi}}
\newcommand{\ochb}{\dagg{\oscgreek{\chi}}}
\newcommand{\NN}{\mathbf{N}}
\newcommand{\ccc}{\mathbf{C}}
\newcommand{\es}{{\varnothing}}
\newcommand{\fs}{{\overline{\varnothing}}}
\newcommand{\dt}[1]{\bar{#1}}
\newcommand{\udt}[1]{{#1}}

\newcommand{\qlax}{\mathcal{R}}
\newcommand{\flax}{\mathcal{L}}
\newcommand{\plax}{L}
\newcommand{\qop}{\mathbf{Q}}

\renewcommand{\Re}{\mathrm{Re}}

\newmuskip\pFqmuskip
\newcommand*\pFq[6][8]{%
    \begingroup
    \pFqmuskip=#1mu\relax
    \mathcode`\,=\string"8000
    \begingroup\lccode`\~=`\,
    \lowercase{\endgroup\let~}\pFqcomma
    {}_{#2}F_{#3}{\left(\genfrac..{0pt}{}{#4}{#5};#6\right)}%
    \endgroup}
\newcommand{\pFqcomma}{\mskip\pFqmuskip}

\newcommand{\mt}{{\tilde{m}}}
\newcommand{\mtb}{{\tilde{\mathbf{m}}}}
\newcommand{\mb}{{\mathbf{m}}}
\newcommand{\mmz}{\hat{m}}
\newcommand{\middlepart}{\mathbb{M}}

\def\be{\begin{eqnarray}}
\def\ee{\end{eqnarray}}
\def\no{\nonumber}

\newcommand{\eqndot}{\; .}
\newcommand{\eqncom}{\; ,}

\newcommand{\dd}{\mathrm{d}}
\newcommand{\nfsym}{$\mathcal{N}\! = \! 4$ SYM\xspace}

\DeclareMathOperator{\str}{str}
\DeclareMathOperator{\tr}{tr}

\newcommand{\bra}[1]{\langle #1 |}
\newcommand{\ket}[1]{| #1 \rangle}

\newcommand{\abs}[1]{{ | {#1} | }}
\newcommand{\gr}[1]{{ | {#1} | }}
\newcommand{\sfrac}[2]{{\textstyle\frac{#1}{#2}}}

\newcommand{\tw}{\tau}

\DeclareRobustCommand{\eulerian}{\genfrac<>{0pt}{}}

\newcommand{\HL}{\Phi}
\newcommand{\zop}{\mathcal{Z}}

\makeatletter
\g@addto@macro\bfseries{\boldmath}
\makeatother

\makeatletter
\newcases{rrcases}{\quad}{%
  \hfil$\m@th\displaystyle{##}$}{$\m@th\displaystyle{##}$\hfil}{\lbrace}{.}
\makeatother

\begin{document}

\begingroup\parindent0pt
\begin{flushright}\footnotesize
\texttt{HU-MATH-2017-04}\\
\texttt{HU-EP-17/14}\\
\texttt{TCDMATH-17-13}\\
\end{flushright}
\vspace*{3em}
\centering
\begingroup\LARGE
\bf
%Bootstrapping the operatorial Q-system of\\ 
%non-compact super spin chains
Evaluation of the operatorial Q-system\\ 
for non-compact super spin chains
\par\endgroup
\vspace{3.5em}
\begingroup\large
{\bf Rouven Frassek}$\,^a$, 
{\bf Christian Marboe}$\,^b$, 
{\bf David Meidinger}$\,^c$ 
\par\endgroup
\vspace{2em}
\begingroup\sffamily\footnotesize
$^a\,$Institut des Hautes Études Scientifiques,\\
   35 Route De Chartres, 91440 Bures-sur-Yvette, France\\
frassek@ihes.fr\\
\vspace{1em}
$^b\,$School of Mathematics,
    Trinity College Dublin, \\
College Green, Dublin 2, Ireland\\
marboec@tcd.ie\\
\vspace{1em}
$^c\,$Institut für Mathematik und Institut für Physik,
Humboldt-Universität zu Berlin,\\
IRIS Gebäude,
Zum Großen Windkanal 6,
12489 Berlin, Germany\\
 david.meidinger@physik.hu-berlin.de
\par\endgroup
\vspace{5em}

\vfill
\begin{abstract}
\noindent
We present an approach to evaluate the full operatorial Q-system of all 
$\mathfrak{u}(p,q|r+s)$-invariant spin chains with representations 
of Jordan-Schwinger type. 
In particular, this includes the super spin chain of planar $\mathcal{N}=4$ 
super Yang-Mills theory at one loop in the presence of a diagonal twist.
Our method is based on the oscillator construction of Q-operators.
The Q-operators are built as traces over Lax operators which are 
degenerate solutions of the Yang-Baxter equation.
For non-compact representations these Lax operators may contain multiple infinite 
sums that conceal the form of the resulting functions.
We determine these infinite sums and calculate the matrix elements of the lowest level Q-operators. 
Transforming the Lax operators corresponding to the Q-operators into a representation 
involving only finite sums allows us to take the supertrace and to obtain the
explicit form of the Q-operators in terms of finite matrices for a given magnon sector.
Imposing the functional relations, we then bootstrap
the other Q-operators from those of the lowest level.
We exemplify this approach for non-compact spin~$-s$ spin chains
and apply it to \nfsym at the
one-loop level using the BMN vacuum as an example.
\end{abstract}

\endgroup

\thispagestyle{empty}

\newpage
\tableofcontents
\newpage

%%%%%%%%%%%%%%%%%%%%%%%%%%%%%%%%%%%%%%%%%%%%%%%%%%%%%%%%%%%%%%%%%%%%%%%%%%%%%%%
\section{Introduction}
%%%%%%%%%%%%%%%%%%%%%%%%%%%%%%%%%%%%%%%%%%%%%%%%%%%%%%%%%%%%%%%%%%%%%%%%%%%%%%%

Non-compact spin chains are quantum  integrable models that appear in certain limits of four-dimensional quantum field theories \cite{Lipatov:1993yb,Beisert:2010jr,Nekrasov:2009rc,Dorey:2011pa}.
In contrast to their compact counterparts the physical or quantum space of non-compact spin chains is infinite-dimensional.
Though physically these spin chains are of very different nature, they can be uniformly described  algebraically in the framework of the quantum inverse scattering method, see e.g.  \cite{Faddeev:1996iy}. 
Here a commuting family of transfer matrices is built from so called R-matrices which are solutions to the Yang-Baxter equation which are studied systematically in the theory of the Yangian and the universal R-matrix \cite{Jimbo:1985zk,Drinfeld:1985rx}. 
For a given quantum space, transfer matrices are constructed from the monodromy of R-operators by tracing over the auxiliary space. Due to the Yang-Baxter equation the transfer matrices built in this way commute for different representations in the auxiliary space.
They satisfy certain functional equations which arise from fusion in the auxiliary space, see e.g. \cite{Zabrodin:1996vm} for an overview.

A distinguished role among the commuting operators is taken by the Q-operators \cite{Baxter:1972hz}. They are related to the transfer matrices via the quantum Wronskian and satisfy so called QQ-relations. For the case of interest see \cite{Tsuboi:2009ud} and references therein where these relations are discussed on the level of eigenvalues, i.e. Q-functions.
In the pioneering work \cite{Bazhanov:1994ft,Bazhanov:1996dr,Bazhanov:1998dq},
Q-operators were constructed as traces over an infinite-dimensional oscillator space. We refer the reader to the more recent works \cite{hernandez2012asymptotic,frenkel2015baxter,boos2016oscillator,boos2017oscillator} for a mathematical discussion of these infinite-dimensional representations.
Similarly, Q-operators for spin chains with diagonal twist can be constructed in the framework of the quantum inverse scattering method.
The corresponding Lax operators for 
certain compact q-deformed higher rank spin chains were written down explicitly in \cite{Bazhanov:2008yc}
and derived from the universal R-matrix in \cite{Boos:2010ss} 
, see also \cite{Antonov:1996ag,Rossi:2002ed,korff2004auxiliary} for earlier works. 
In the rational case the relevant Lax operators were obtained in a series of papers \cite{Bazhanov:2010ts,Bazhanov:2010jq,Frassek:2010ga,Frassek:2011aa,Frassek:xxx} for $\mathfrak{gl}(N|M)$.
These solutions allow for the definition of Q-operators for more general representations and in particular as discussed in this article for representations of the non-compact super algebras $\mathfrak{u}(p,q|r+s)$
acting on the quantum space.  

While for compact spin chains the Lax operators derived in  \cite{Frassek:2011aa}  can straightforwardly be used to evaluate the matrix elements of Q-operators, it is much more involved to extract their matrix elements in the case of non-compact representations. 
The first obvious reason is that the quantum space is infinite-dimensional, however noting that the Q-operators are block diagonal this problem can be overcome by considering magnon blocks separately. 
The other, more serious issue arises because the Lax operators relevant to construct Q-operators were derived in the form of a Gauss decomposition. Besides its beauty, this form is rather inconvenient for practical purposes. One has to sum over intermediate states to compute explicit matrix elements,
and in the case of non-compact representations there are potentially infinitely many such states. 

In this paper we overcome these difficulties which arise when evaluating Q-operators for non-compact spin chains of Jordan-Schwinger type and present an efficient method to determine the full operatorial Q-system%
\footnote{
We use the term Q-system to denote the full set of Q-operators or Q-functions of a given integrable model together with their functional relation. The term is also used for the system of functional relations among characters of Kirillov-Reshetikhin modules, see \cite{Kirillov1990}.
}
for a fixed magnon block. 
Section~\ref{sec:qoperatorsjs} is a brief review where we introduce the operatorial Q-system and the corresponding Lax operators, and discuss the Jordan-Schwinger type representations in the quantum space.
We then first discuss non-compact spin~$-s$ chains in Section~\ref{sec:sl2}. In this case there are two non-trivial Lax operators, one involving an infinite sum, whose matrix elements we evaluate in full detail. 
We generalise our approach to higher rank super spin chains in Section~\ref{sec:finiterep}. Here we discuss which Q-operators involve infinite sums and provide a decomposition of the Lax operators on which our approach is based.
For the lowest level Lax operators we present an integral formula which conveniently allows to evaluate them in terms of rational functions. The evaluation of higher level Lax operators is discussed in Section~\ref{sec:higher}.
In Section~\ref{sec:qsys}, we use the integral representation of the lowest level Lax operators to calculate the matrix
elements of the corresponding Q-operators as finite matrices in each
magnon block. We furthermore show how the remaining Q-operators can 
efficiently be determined from this data.
In Section~\ref{sec:vacuum} we show how to apply these methods to the 
\nfsym spin chain in the presence of a full diagonal 
twist, and calculate the Q-functions of the BMN vacuum of arbitrary length.
We conclude our work in Section~\ref{sec:conclusion} and speculate about the application of Q-operators to the Quantum Spectral Curve of \nfsym \cite{Gromov:2013pga,Gromov:2014caa}. 
We provide further information on the operatorial Q-systems in the appendix.
In particular, to facilitate the application of our results, we collect all 
formulas which are needed for the calculation of the Q-systems in
Appendix~\ref{sec:formulas}. These include formulas for
the matrix elements of the Lax operators, and for the evaluation 
of the supertraces over the auxiliary Fock spaces.

%%%%%%%%%%%%%%%%%%%%%%%%%%%%%%%%%%%%%%%%%%%%%%%%%%%%%%%%%%%%%%%%%%%%%%%%%%%%%%%
%%%%%%%%%%%%%%%%%%%%%%%%%%%%%%%%%%%%%%%%%%%%%%%%%%%%%%%%%%%%%%%%%%%%%%%%%%%%%%%
%%%%%%%%%%%%%%%%%%%%%%%%%%%%%%%%%%%%%%%%%%%%%%%%%%%%%%%%%%%%%%%%%%%%%%%%%%%%%%%
\section{Q-operators for representations of oscillator type}
\label{sec:qoperatorsjs}
%%%%%%%%%%%%%%%%%%%%%%%%%%%%%%%%%%%%%%%%%%%%%%%%%%%%%%%%%%%%%%%%%%%%%%%%%%%%%%%

In this section we present the derivation of the Lax operators ($\qlax$-operators) which allow to construct 
the Q-operators of $\mathfrak{gl}(N|M)$ spin chains with representations 
realised via Jordan-Schwinger oscillators as traces of monodromy matrices.
The general construction reviewed here was developed in a series of papers:
The derivation of the $\qlax$-operators follows the bosonic case in~\cite{Frassek:2011aa}
but incorporates the supersymmetric Lax matrices derived 
in~\cite{Frassek:2010ga}. For the Lax operators of bosonic models, Schwinger oscillators were 
discussed in \cite{Meneghelli:thesis}.
The more general derivation of the Lax operators for generalised rectangular 
representations is unpublished~\cite{Frassek:xxx} while expressions for the 
resulting operators can be found in~\cite{Frassek:thesis}. 

\subsection{Lax operators for Q-operators}\label{sec:laxe}

The study of supersymmetric rational spin chains goes back to 
Kulish~\cite{Kulish:1985bj} who introduced the $\mathfrak{gl}(N|M)$~invariant 
Lax operators
\begin{equation}
    \flax(z)=z+\sum_{a,b=1}^{N+M}(-1)^{\gr{b}} e_{ab}E_{ba}\,,
    \label{eq:slax}
\end{equation} 
intertwining arbitrary representations of $\mathfrak{gl}(N|M)$ with the defining fundamental one.
Here the indices take the values $a,b=1,\ldots,N+M$ while $\gr{a}$ denotes the grading $\gr{\text{fermion}}=1$ and 
$\gr{\text{boson}}=0$. The $\mathfrak{gl}(N|M)$  generators $E_{ab}$ satisfy the  commutation relations
\begin{equation}
    [E_{ab},E_{cd}] = 
    \delta_{bc}E_{ad}-(-1)^{(\gr{a}+\gr{b})(\gr{c}+\gr{d})}\delta_{da}E_{cb}\,,
    \label{eq:glnm}
\end{equation} 
where we defined the graded commutator as $ [X,Y]=XY-(-1)^{\gr{X}\gr{Y}}YX$.
The generators $e_{ab}$ in \eqref{eq:slax} denote 
the defining fundamental generators of $\mathfrak{gl}(N|M)$  satisfying $e_{ab}e_{cd}=\delta_{bc}e_{ad}$. In the following we restrict to the Schwinger oscillator realisation
\begin{equation}
    E_{ab}=\ochb_{a}\och_{b}\,,
    \label{eq:schwinger}
\end{equation} 
where $[\och_{a},\ochb_{b}] =\delta_{ab}$.

The Lax operators for Q-operators with the defining representation of 
$\mathfrak{gl}(N|M)$ at each spin chain site (the so-called quantum space) 
were derived in \cite{Frassek:2010ga}, and are given by
\begin{equation}
    \plax_I(z) =
    \left(\begin{array}{cc}
        (z-s_I)\delta_{ab}-(-1)^{\gr{b}}\oxb_{\udt a \dt a}\ox_{\dt a \udt b} & 
        \oxb_{\udt a \dt b} \\
        -(-1)^{\gr{b}}\ox_{\dt a \udt b} &
        \delta_{\dt a \dt b}
    \end{array}\right)\,.
    \label{eq:splax}
\end{equation} 
There are $2^{N+M}$ such Lax operators labelled by the set $I\subseteq \{1,\ldots,M+N\}$. 
The notation here and in the rest of this article is as follows:
we sum over repeated indices (appearing two or more times);
unbarred indices take values $\udt a, \udt b\in I$ while barred ones 
take values in its complement, $\dt a,\dt b\in \bar I$.
The $(N|M)\times(N|M)$ matrix in \eqref{eq:splax} is written in terms of the sub blocks under
this decomposition.%
\footnote{We remark that quantities labelled by the set $I$ depend on the partition $I\cup\bar I=\{1,\cdots,N+M\}$. We leave the dependence on this full set implicit.}
The shift $s_I$ in the spectral parameter $z$ is introduced for convenience and reads 
 \begin{equation}
    s_I=\frac{\sum_{\dt a\in \bar I}(-1)^{\gr{\dt a}}}{2}\,.
    \label{eq:def_shift}
\end{equation}
The oscillators 
$(\ox_{\dt a \udt a} ,\oxb_{\udt a \dt a})$ satisfy the graded Heisenberg algebra
\begin{equation}
    [\ox_{\dt a \udt a},\oxb_{\udt b \dt b}] = 
    \ox_{\dt a \udt a}\oxb_{\udt b \dt b} 
    - (-1)^{(\gr{\udt a}+\gr{\dt a})(\gr{\udt b}+\gr{\dt b})}
    \oxb_{\udt b \dt b}\ox_{\dt a \udt a}
    = \delta_{\udt a \udt b}\delta_{\dt a \dt b}\,.
    \label{eq:comosc}
\end{equation} 
We can write down the defining Yang-Baxter equation for the 
$\qlax$-operators which are the building blocks for Q-operators when the sites of the quantum space
are in a representation space different from the fundamental representation. As in the bosonic case \cite{Frassek:2011aa} this relation is given by
\begin{equation}
    \flax(x-y)\plax_{I}(x)\qlax_{I}(y) = 
    \qlax_{I}(y)\plax_{I}(x)\flax(x-y)\,.
    \label{eq:yberll}
\end{equation} 
The form of $\qlax$-operators was obtained in \cite{Frassek:xxx} and spelled out in \cite{Frassek:thesis}. The derivation follows \cite{Frassek:2011aa} and, as we will discuss in the following, simplifies significantly in the case which we are interested in.  As for $\mathfrak{gl}(N)$ one takes
a factorised ansatz,
\begin{equation}
    \qlax_{I}(z) = 
    e^{(-1)^{\gr{\udt c}+\gr{\udt c}\gr{\dt c}}
        \oxb_{\udt c \dt c}E_{\udt c \dt c}}
        \,
    \qlax_{0}^{I}(z)
    \,
    e^{-(-1)^{\gr{\udt d}\gr{\dt d}+\gr{\udt d}+\gr{\dt d}}
        \ox_{\dt d \udt d}E_{\dt d \udt d}}\,,
        \label{eq:laxus}
\end{equation} 
and ends up with a difference equation for the middle part $ \qlax_{0}^{I}(z)$. The solution to the difference equation simplifies significantly for the choice of generators \eqref{eq:schwinger}.
For representations of this type one finds that $\qlax_{0}^{I}(z)$ can be written in terms of a single 
Gamma~function
\begin{equation}
    \qlax_{0}^{I}(z) = \rho_{I}(z)\, 
    \Gamma\left(z+1-s_I-\textstyle\sum_{\dt a}E_{\dt a\dt a}\right)
    \eqndot
    \label{eq:r0}
\end{equation} 
Here $\rho_I$ denotes a normalisation not fixed by the 
Yang-Baxter equation \eqref{eq:yberll}. As we will see, a good choice for it
is given by
 \begin{equation}
    \rho_I(z)=\frac{1}{\Gamma(z+1-s_I-\ccc)}
    \eqncom
    \label{eq:norm}
\end{equation}
which depends on the  central charge $\ccc$ that can be expressed in terms of the number operators $\NN_a=\ochb_{a}\och_{a}$ as 
\begin{equation}
    \ccc=\sum_{a=1}^{N+M} \NN_a
    \eqndot
    \label{eq:def_central_charge}
\end{equation}
We conclude that
\begin{equation}
    \qlax_{I}(z) =  
    e^{(-1)^{\gr{\udt c}+\gr{\udt c}\gr{\dt c}}
        \oxb_{\udt c \dt c}\ochb_{\udt c}\och_{\dt c}}
        \,
    \frac{
        \Gamma(z+1-s_I-\ochb_{\dt a}\och_{\dt a})
    }{
        \Gamma(z+1-s_I-\ccc)
    }
    \,
    e^{-(-1)^{\gr{\udt d}\gr{\dt d}+\gr{\udt d}+\gr{\dt d}}
        \ox_{\dt d \udt d}\ochb_{\dt d}\och_{\udt d}}
        \label{eq:fullax}
\end{equation} 
solves the Yang-Baxter equation in \eqref{eq:yberll}. 
The normalisation \eqref{eq:norm} ensures that for the empty set
$ \qlax_{\es}(z)=1$ 
and renders  $ \qlax_{I}(z)$ a polynomial in $z$ for compact 
representations. 
Finally we note that the middle part of \eqref{eq:fullax} involves multiple Gamma functions for more general representations, see~\cite{Frassek:2011aa,Frassek:xxx,Frassek:thesis}. However, for Jordan-Schwinger type representations only one Gamma function appears, cf.~\cite{Meneghelli:thesis}.

\subsection{Definition of the Q-operators}
\label{sec:qs}

Using the Lax operators $\plax_I(z)$ defined in \eqref{eq:splax},
the Q-operators for $\mathfrak{gl}(N|M)$ rational spin chains were introduced in  \cite{Frassek:2010ga} for the defining fundamental representations at each site of the quantum space.
We are interested in more general representations of oscillators type, cf. \eqref{eq:schwinger}.
However, as discussed in e.g. \cite{Frassek:2011aa} for $\mathfrak{gl}(N)$ the construction of the Q-operators and the functional relations among them should be independent of the quantum space. Thus, following \cite{Frassek:2010ga} we define the Q-operators as
\begin{equation}
 \qop_I(z)=e^{iz\sum_{a\in I}(-1)^{|a|}\phi_a}\,\widehat\str\,  \mathcal{M}_I(z)\,.
 \label{eq:qop}
\end{equation} 
Here the monodromy $\mathcal{M}_I$ is built from the tensor product of the $\qlax$-operators in  \eqref{eq:fullax} in the space of oscillators $(\och,\ochb)$ and multiplication in the auxiliary space of oscillators $(\ox,\oxb)$ as
\begin{equation}
 \mathcal{M}_I(z)=\qlax_{I}^{[1]}(z)\otimes \qlax_{I}^{[2]}(z)\otimes\ldots\otimes \qlax_{I}^{[L]}(z)\,.
\end{equation} 
The normalised supertrace $\widehat\str$ is defined by 
\begin{equation}
 \widehat\str\, X =\frac{\str e^{-i\sum_{a,b}(\phi_a-\phi_b)\NN_{ab}} X}{\str e^{-i\sum_{a,b}(\phi_a-\phi_b)\NN_{ab}}}\,,
 \label{eq:str}
\end{equation} 
where $\str$ denote the ordinary supertrace over the auxiliary Fock space
spanned by the states generated from acting with the operators $\oxb_{\udt a \dt a}$ on a Fock vacuum satisfying $\ox_{\dt a \udt a}\ket{0}=0$. These states are labelled by the values of the number operators 
\begin{equation}
 \NN_{ab}=\oxb_{ab}\ox_{ba}\,,
\end{equation} 
where no sum is implied over the indices $a$ and $b$.
The twist parameters $\phi_a$ which can be interpreted as
Aharonov-Bohm phases, cf. \cite{Frassek:2010ga},
break the $\mathfrak{gl}(N|M)$ invariance down to its diagonal subalgebra.
They are required for the convergence of the supertraces.
Note that a regularisation is needed to make some of the traces converge, even 
in the presence of twists;
one can use an $i\varepsilon$ prescription for the twists, such that 
$\Re(\exp({-i\sum_{a,b}(\phi_a-\phi_b)}))< 1$, see~\cite{Bazhanov:2010jq}.

The Q-operators defined in \eqref{eq:qop}  commute with each other, $[\qop_I(z),\qop_{I'}(z')]=0$,
and, as a consequence of the Yang-Baxter equation \eqref{eq:yberll}, also with the transfer matrix built from the Lax operators $\mathcal{L}$ realised via \eqref{eq:schwinger}.
For a discussion on how to obtain the Hamiltonian from the Q-operators we refer the reader to \cite{Frassek:2012mg}.
Further it was argued in \cite{Frassek:2010ga} that depending on the grading the Q-operators satisfy either the bosonic QQ-relations
\begin{equation}
    \Delta_{ab} \qop_{I\cup \{a,b\}}(z) \qop_I(z)= \qop_{I\cup \{a\}}(z+\sfrac{1}{2}) \qop_{I\cup \{b\}}(z-\sfrac{1}{2})-\qop_{I\cup\{a\}}(z-\sfrac{1}{2}) \qop_{I\cup \{b\}}(z+\sfrac{1}{2})\,, \label{eq:QQb}
\end{equation}
where $|a|=|b|$ or the fermionic QQ-relations
\begin{equation}
    \Delta_{ab} \qop_{I\cup \{a\}}(z) \qop_{I\cup \{b\}}(z)= \qop_{I\cup\{a,b\}}(z+\sfrac{1}{2}) \qop_{I}(z-\sfrac{1}{2})-\qop_{I\cup \{a,b\}}(z-\sfrac{1}{2}) \qop_{I}(z+\sfrac{1}{2})\,, \label{eq:QQf}
\end{equation} 
where $|a|\neq|b|$. Here we defined the trigonometric prefactor
\begin{equation}
    \Delta_{ab}= (-1)^{\gr{a}} 2i\sin\left(\frac{\phi_a-\phi_b}{2}\right)\,.
 \label{eq:delta}
\end{equation} 
The set of all Q-operators can be visualised on a hypercubic Hasse diagram,
representing the partial order induced by the inclusion of indices, see for example
\cite{Tsuboi:2009ud}.
The relations \eqref{eq:QQb} and \eqref{eq:QQf} then constrain the operators
on each quadrilateral of this diagram. 
It is straightforward to compute the Q-operators for the empty set  $I=\es$ and the full set $I=\{1,\ldots,N+M\}=\fs$. Using the normalisation in \eqref{eq:norm} one finds
\begin{equation}
    \qop_\es(z) = 1\,,
    \qquad
    \qop_\fs(z) = \left(\frac{\Gamma(z+1)}{\Gamma(z+1-\ccc)}\right)^L\,,
    \label{eq:qemptyfull}
\end{equation}
where we imposed the constraint%
\footnote{
    For $\mathfrak{gl}(N|N)$ one has the additional constraint $ \sum_{a=1}^{2N}\phi_a=0$.
}
\begin{equation}
    \sum_{a=1}^{N+M}(-1)^{\gr{a}}\phi_a=0
    \label{eq:twistconstraint}
    \eqndot
\end{equation}
This relation is needed for $\qop_{\fs}$ to be a rational function of the
spectral parameter.

\subsection{Representation in the quantum space}
\label{sec:representation}

So far we did not specify a representation in the quantum space.
Parts of our derivations will be independent of the concrete representation,
but calculations of explicit matrix elements of course require a concrete 
knowledge of the representation space.
Here we focus on unitary highest- or lowest-weight representations of $\mathfrak{u}(p,q|r+s)$
of oscillator type which were first investigated in \cite{Bars:1982ep}.
To specialise to a real form of the algebra, 
we have to indicate, in addition to the grading $\gr{a}=0,1$,
which directions have opposite sign under conjugation that can be realised via
a particle-hole transformation. We indicate these using the variables
\begin{equation}
    \omega_{a} = 
    \begin{cases}
        +1 &
        \text{if oscillator $a$ is not transformed} \\
        -1&
        \text{if oscillator $a$ is transformed} 
    \end{cases}
    \eqndot
    \label{eq:phindicator}
\end{equation}
Then the generators $E_{ab}=\ochb_a\och_b$ can be realised by the oscillators
\begin{equation}
    (\och_a,\ochb_a)=\begin{rrcases}
        \;\;\;
        (\oa_a,\oab_a)
        &\qquad\text{for }\gr{a}=0\text{ and }\omega_a=+1\\
        (\obb_a,-\ob_a)
        &\qquad\text{for }\gr{a}=0\text{ and }\omega_a=-1\\
        (\oc_a,\ocb_a)
        &\qquad\text{for }\gr{a}=1\text{ and }\omega_a=+1\\
        (\odb_a,\od_a)
        &\qquad\text{for }\gr{a}=1\text{ and }\omega_a=-1
    \end{rrcases}
    \eqndot
    \label{eq:realoscs}
\end{equation}
These oscillators act on a Fock space with 
a vacuum state satisfying
$
\oa_a\ket{0}=
\ob_a\ket{0}=
\oc_a\ket{0}=
\od_a\ket{0}=0
$, such that an orthonormal basis is given by
\begin{equation}
    \ket{\mb}
    =
    \ket{m_1,\ldots,m_K}
    =
    \frac{\left\{\begin{smallmatrix}\oab_1\\\obb_1\\\ocb_1\\\odb_1\end{smallmatrix}\right\}^{m_1}}{\sqrt{m_1!}}
    \cdots
    \frac{\left\{
        \begin{smallmatrix}\oab_{K}\\\obb_{K}\\\ocb_{K}\\\odb_{K}\end{smallmatrix}\right\}^{m_{K}}}{\sqrt{m_{K}!}}
    \ket{0,\ldots,0}
    \eqncom
    \label{eq:defstates}
\end{equation}
with $K=N+M=p+q+r+s$.
Since the oscillators obey the standard conjugation
$\oa^\dagger=\oab$,
$\ob^\dagger=\obb$,
$\oc^\dagger=\ocb$ and
$\od^\dagger=\odb$,
the generators are those of $\mathfrak{u}(p,q|r+s)$, satisfying
\begin{equation}
    E_{ab}^\dagger = 
    \omega_a^{1+\gr{a}}
    \omega_b^{1+\gr{b}}
    E_{ba}
    \eqndot
\end{equation}

Finally, the Fock space contains a series of representations labelled
by the central charge $\ccc=\sum_a\NN_a$ where the number operators have to
be expressed in terms of oscillators $\oa$, $\ob$, $\oc$ and $\od$ via \eqref{eq:realoscs}.
These representations are of highest or lowest weight type depending on the order
of the different types of oscillators.
\section{Non-compact Heisenberg spin chains}
\label{sec:sl2}

In this section we provide the formulas necessary to evaluate Q-operators of non-compact 
Heisenberg spin chains, which include for example the spin $-1$
model which is of interest for QCD in the Regge limit \cite{Faddeev:1994zg}.
These models constitute the simplest non-trivial case which can be treated in the more general framework presented in Section~\ref{sec:finiterep} and \ref{sec:qsys}. 

While the formula for the Lax operators \eqref{eq:fullax} is extremely compact,
it is rather inconvenient for practical purposes where the matrix elements of the Lax operators and Q-operators are of interest. In particular, for non-compact representations we encounter infinite sums. 
To understand this problem we consider the $\qlax$-operators \eqref{eq:fullax} with $|I|=1$
which for the case of $\mathfrak{gl}(2)$ are given by
\begin{equation}
   \qlax_{\{a\}}(z) =  
    e^{\oxb_{\udt a \dt a}E_{\udt a\dt a}}
    \frac{
        \Gamma(z+\frac{1}{2}-E_{\dt a\dt a})
    }{
        \Gamma(z+\frac{1}{2}-\ccc)
    }
    e^{-
        \ox_{\dt a \udt a}E_{\dt a\udt a}}\,,
        \label{eq:fullaxsl2}
\end{equation} 
where $a=1,2$ and $\dt a \neq a$. For infinite-dimensional representations the Lax operator contains two infinite sums emerging from the exponential functions. Using the algebraic relations in \eqref{eq:glnm} we note that the Lax operators can be rewritten as
\begin{equation}
    \qlax_{\{a\}}(z) = \sum_{n=-\infty}^{+\infty} 
 (\oxb_{\udt a \dt a}E_{\udt a\dt a})^{\theta(+n)|n|}
    \mathbb{M}_{\{a\}}(z;|n|)
       (-
        \ox_{\dt a \udt a}E_{\dt a\udt a})^{\theta(-n)|n|}\,,
       \label{eq:fullaxsl22}
\end{equation} 
with $\theta(-m)=\theta(0)=0$ and $\theta(m)=1$ for $m\in\mathbb{N}_+$. The middle part is given by an infinite sum and only depends on Cartan elements 
\begin{equation}
  \mathbb{M}_{\{a\}}(z;|n|)=\frac{1}{|n|!}\frac{\Gamma(z+\frac{1}{2}-E_{\dt a\dt a})}{ \Gamma(z+\frac{1}{2}-\ccc)}\,_3F_2(E_{\dt a\dt a}-\lambda_1,E_{\dt a\dt a}-\lambda_2+1,-\NN_{a\dt a};1+|n|,E_{\dt a\dt a}+\frac{1}{2}-z;1)\,,
  \label{eq:laxmid}
\end{equation} 
with the $\mathfrak{gl}(2)$ weights $\lambda_1$ and $\lambda_2$.\footnote{Note that in the rank $1$ case the reformulation \eqref{eq:fullaxsl22} with \eqref{eq:laxmid} is also valid for representations that are not of Jordan-Schwinger type.}

We are interested in highest-weight state representations of the type discussed in Section~\ref{sec:representation}. To describe  non-compact spin chains with spin~$-s$, where $s$ is a positive half integer,  we take the Jordan-Schwinger realisation \eqref{eq:schwinger} and perform a particle hole 
transformation on the oscillators of type $1$:
\begin{equation}
 (\ochb_1,\och_1)\rightarrow  (-\ob,\obb)  \,,\qquad  (\ochb_2,\och_2)\rightarrow (\oab,\oa)\,.
\end{equation} 
For convenience we use a notation different from the rest of this article and label the states in the spin $-s$ representation
as
\begin{equation}
 |m\rangle_s=|2s-1+m,m\rangle\,,
 \label{eq:statesrank1}
\end{equation} 
cf.~\eqref{eq:defstates}. 
The highest-weight state $ |0\rangle_s$ then satisfies
\begin{equation}
 E_{12}|0\rangle_s=0\,,\qquad  E_{11}|0\rangle_s=\lambda_1|0\rangle_s=-2s|0\rangle_s\,,\qquad  E_{22}|0\rangle_s=\lambda_2|0\rangle_s = 0
 \eqncom
\end{equation} 
and the other states of the representation can be generated from $|0\rangle_s$ by acting with the operator $E_{21}$. The central charge takes the value $\ccc|m\rangle_s=-2s|m\rangle_s$.
%The representation is characterised by the $\mathfrak{gl}(2)$ weights $\lambda_1=-2s$ and $\lambda_2=0$, and the central charge takes the value $\ccc|m\rangle_s=-2s|m\rangle_s$.

Our goal is to obtain the matrix elements of the $\qlax$-operators in \eqref{eq:fullaxsl2}. From \eqref{eq:qemptyfull} we find that $\qlax_\es(z)=1$ and $\qlax_\fs(z)=\frac{1}{(z+1)_{2s}}$,
where the Pochhammer symbol is defined by $(a)_n\coloneqq \Gamma(a+n)/\Gamma(a)$.%
\footnote{In the following we will sometimes consider Pochhammer symbols where $a$ can be a negative integer. In this case we define the symbol using the identity
\(
\frac{\Gamma(a+n)}{\Gamma(a)}
=(-1)^n
\frac{\Gamma(1-a)}{\Gamma(1-a-n)}
\), which follows from Euler's reflection formula.}

It is rather straightforward to obtain the matrix elements of $\qlax_{\{2\}}$. They are polynomials in the spectral parameter $z$ and can be obtained noting that the series representation of the hypergeometric function in \eqref{eq:laxmid} truncates. One finds
\begin{equation}
\begin{split}
    _s\langle \tilde m|\qlax_{ \{2\} }(z)| m\rangle_s&=\sqrt{\frac{\max(m,\tilde m)!}{\min(m,\tilde m)!}\frac{\max(2s-1+m,2s-1+\tilde m)!}{\min(2s-1+m,2s-1+\tilde m)!}}\\[5pt]
    &\;\times\oxb_{21}^{\,\theta(\tilde m- m)|\tilde m- m|}\; \middlepart_{\{2\}}(z,\NN_{21},|m-\tilde m|,\min(m,\tilde m))\;\ox_{12}^{\,\theta( m-\tilde m)|m-\tilde m|}\,,
    \end{split}
    \label{eq:formR2}
\end{equation} 
with the middle part which is diagonal in the auxiliary space
\begin{equation}
    \middlepart_{ \{2\}}(z,\NN_{21},k,l)=(2s-1+l)!\sum_{p=0}^{l}\binom{l}{p}\frac{(\NN_{21}+1+p-l)_{l-p}(z+\frac{1}{2}+2s)_{p}}{(2s-1+p)!(k+l-p)!}
    \eqndot
    \label{eq:middleR2}
\end{equation}
Here $\theta(-m)=\theta(0)=0$ and $\theta(m)=1$ for $m\in\mathbb{N}_+$.

However, as already noted in \cite{Frassek:2011aa} the operator $\qlax_{\{1\}}$ yields infinite sums when evaluated naively, since there are only raising operators acting on the states, cf.~\eqref{eq:fullaxsl2}.
This makes it difficult to evaluate its matrix elements concretely in terms of 
rational functions.
Nevertheless, as we will show in Section~\ref{sec:derivation}, using the integral
representation of the hypergeometric function and the Euler transformation
the matrix elements can be obtained from \eqref{eq:fullaxsl22} and written as
\begin{equation}
    \begin{aligned}
        &_s\langle \tilde m|\qlax_{\{1\}}(z)| m\rangle_s=\sqrt{\frac{\max(m,\tilde m)!}{\min(m,\tilde m)!}\frac{\max(2s-1+m,2s-1+\tilde m)!}{\min(2s-1+m,2s-1+\tilde m)!}}
    \\[6pt]
    &\qquad\times
    (-\oxb_{12})^{\,\theta(m-\tilde m)|m-\tilde m|} \middlepart_{ \{1\}}(z,\NN_{12},|m-\tilde m|,\max(m,\tilde m))(-\ox_{21})^{\,\theta(\tilde m - m)|\tilde m -m|}
    \eqncom
    \label{eq:formR1}
    \end{aligned}
\end{equation} 
with the middle part taking the simple form
\begin{equation}
    \middlepart_{ \{1\}}(z,\NN_{12},k,l)=\frac{ \middlepart_{\{2\}}(z,\NN_{12},k,l-k)}{(z-\NN_{12}-l+\frac{1}{2})_{2l-k+2s}}\,.
    \label{eq:middleR1}
\end{equation} 
We see that also this non-polynomial $\qlax$-operator can be written in a very compact form and observe that the
resulting expression is very similar to the polynomial $\qlax$-operator. In particular, both are simple
rational functions of the spectral parameter and the auxiliary oscillators. 
The only difference is the dependence on the
auxiliary space number operators in the denominator.
This has important consequences for the analytic structure of the resulting Q-operators.

The matrix elements of the corresponding Q-operators \eqref{eq:qop} can now be derived as
\begin{equation}
    _s\langle\tilde{\mathbf{m}}|\qop_{\{a\}}(z)|{\mathbf{m}}\rangle_s=e^{iz\phi_{a}}\,\widehat \tr\,_s   \langle \tilde m_1|\qlax_{\{a\}}(z)| m_1\rangle_s\cdots\,_s  \langle  \tilde m_L|\qlax_{\{a\}}(z)|  m_L\rangle_s\,.
\end{equation} 
The big advantage of first evaluating the $\qlax$-operators in the quantum space and subsequently taking the trace in the auxiliary space is that we can restrict to individual magnon sectors with $M=\sum_{i=1}^L m_i=\sum_{i=1}^L \tilde m_i$. For each such sector, the Q-operators can then 
be realised as matrices of finite size.
As we will show in Section~\ref{sec:qsys}, the matrix elements of the
Q-operators corresponding to the $\qlax$-operators with non-truncating sums are non-rational functions and can be written in terms of the 
Lerch transcendent (Lerch zeta-function) defined as
\begin{equation}
    \label{eq:def_hl}
 \HL^\tw_\ell(z)=\sum_{k=0}^\infty \frac{\tw^{ k}}{(k+z)^\ell}
 \eqndot
\end{equation} 

To give the reader an impression of the resulting Q-functions
we consider the concrete case of a
spin chain with spin $-\frac{1}{2}$.
For small length $L$ and magnon number $M$ the 
Q-operators resulting from the monodromy construction can easily be diagonalised. 
The eigenvalues and eigenvectors 
containing the twist parameters are rather involved. 
For the case $L=2$ and $M=0,1,2$ one easily obtains explicit though rather 
lengthy expressions for the eigenvalues and eigenvectors. 
Due to the constraint \eqref{eq:twistconstraint}, there is only
one independent twist parameter with $\phi_1=-\phi_2$.
For small values of $\phi_1$ the eigenvalues corresponding to the highest-weight states of the untwisted spin chain are given by:
\begin{center}
\begin{tabular}{l|l|l}
    $M$&$Q_{\{1\}}(z)$&$Q_{\{2\}}(z)$\\
   \hline
   0 & $2i\phi_1\big[\psi'(-z-\frac{1}{2})\big]+\mathcal{O}(\phi_1^2)$ & $1$ \\
   1 & $2i\phi_1\times(-4)\times\big[1 + (z+1) \psi'(-z-\frac{1}{2})\big]+\mathcal{O}(\phi_1^2)$ & $(z+1)+\mathcal{O}(\phi_1)$ \\
   2 & $2i\phi_1\times 9 \times\big[(z+1) +(z^2+2z+\sfrac{13}{12}) \psi'(-z-\frac{1}{2})\big]+\mathcal{O}(\phi_1^2)$ & $(z^2+2z+\frac{13}{12})+\mathcal{O}(\phi_1)$ \\
\end{tabular}
\end{center}
Here the non-polynomial Q-functions are expressed in terms of the Polygamma
function $\psi'(z)=\HL^1_2(z)$. 
We observe that for fixed $M$, the prefactors of these functions in $Q_{\{1\}}$ are given by the functions $Q_{\{2\}}$, which are known in closed form and given by Hahn polynomials \cite{Korchemsky:1995be,Eden:2006rx}.
Expanding the factor $\Delta_{12}=2i\phi_1+\mathcal{O}(\phi_1^3)$
in the functional relation \eqref{eq:QQb}, we see that the functions
$\frac{1}{2i\phi_1}Q_{\{1\}}(z)$ and $Q_{\{2\}}(z)$ satisfy the functional relations of the
untwisted spin chain, where the factor $\Delta_{12}$ is not present. 

%%%%%%%%%%%%%%%%%%%%%%%%%%%%%%%%%%%%%%%%%%%%%%%%%%%%%%%%%%%%%%%%%%%%%%%%%%%%%%%
%%%%%%%%%%%%%%%%%%%%%%%%%%%%%%%%%%%%%%%%%%%%%%%%%%%%%%%%%%%%%%%%%%%%%%%%%%%%%%%
%%%%%%%%%%%%%%%%%%%%%%%%%%%%%%%%%%%%%%%%%%%%%%%%%%%%%%%%%%%%%%%%%%%%%%%%%%%%%%%
\section{Non-compact $\qlax$-operators and infinite sums}
\label{sec:finiterep}
%%%%%%%%%%%%%%%%%%%%%%%%%%%%%%%%%%%%%%%%%%%%%%%%%%%%%%%%%%%%%%%%%%%%%%%%%%%%%%%

In Section~\ref{sec:laxe} we introduced the $\qlax$-operators with
Jordan-Schwinger oscillator realisation \eqref{eq:schwinger} in the quantum space. As for more general representations in the quantum space  the $\qlax$-operators naturally decompose into three factors, cf.~\eqref{eq:fullax}. However, as discussed for the case of non-compact spin chains with
$\mathfrak{u}(1,1)$ symmetry in Section~\ref{sec:sl2}, this undoubtedly elegant expression has a drawback when considering non-compact representations. The exponentials appearing on the right (left) in the $\qlax$-operators
\eqref{eq:fullax} do not truncate in the case with only creation (annihilation)
operators in the quantum space.
Thus one has to sum over an infinite tower of states.%
\footnote{Note that at this stage we do not consider the auxiliary space. 
    Later on when evaluating Q-operators we will take the trace over 
    the infinite-dimensional Fock space, see \eqref{eq:qop}.} 
   This issue likewise appears for higher rank algebras $\mathfrak{u}(p,q|r+s)$ as introduced in Section~\ref{sec:representation}.
    Furthermore, in this case even the $\qlax$-operators with truncating sums are complicated,
and naively expanding the exponentials leads to a large number of crossterms which grows exponentially;
most of these terms do not contribute to a given matrix element.

We start with a survey of the Q-system of $\mathfrak{u}(p,q|r+s)$ and discuss its analytic structure. Afterwards we study a different representation of \eqref{eq:fullax}, 
that is simpler to evaluate in the non-compact, but also in the compact, case. We obtain a convenient formula \eqref{eq:middlepart_lowest} to compute the lowest order Lax operators which, as we will see in Section~\ref{sec:qsys}, is sufficient to determine the whole Q-system. Finally we speculate about generalisations beyond the first level.

\subsection{Comments on the Q-system}
\label{sec:com}

\begin{figure}
 \begin{center}

     \begin{tikzpicture}[scale=0.6]
         \draw[] (0,0) -- (5,0);
         \draw[] (0,-1) -- (5,-1);
         \draw[] (0,-3) -- (5,-3);
         \draw[] (0,-4) -- (5,-4);
         \draw[] (0,-5) -- (5,-5);
         \draw[] (0,0) -- (0,-5);
         \draw[] (1,0) -- (1,-5);
         \draw[] (2,0) -- (2,-5);
         \draw[] (4,0) -- (4,-5);
         \draw[] (5,0) -- (5,-5);
         \draw[color=gray] (0,0) -- (2,2);
         \draw[color=gray] (1,0) -- (3,2);
         \draw[color=gray] (2,0) -- (4,2);
         \draw[color=gray] (4,0) -- (6,2);
         \draw[color=gray] (5,0) -- (7,2);
         \draw[color=gray] (2,2) -- (7,2);
         \draw[color=gray] (0+0.5,0+0.5) -- (5+0.5,0+0.5);
         \draw[color=gray] (0+1.5,0+1.5) -- (5+1.5,0+1.5);
         \draw[color=gray] (5,-5) -- (7,-3);
         \draw[color=gray] (5,-4) -- (7,-2);
         \draw[color=gray] (5,-3) -- (7,-1);
         \draw[color=gray] (5,-1) -- (7,1);
         \draw[color=gray] (7,-3) -- (7,2);
         \draw[color=gray] (5+0.5,-5+0.5) -- (5+0.5,0+0.5);
         \draw[color=gray] (5+1.5,-5+1.5) -- (5+1.5,0+1.5);
         \dcirc{0}{0} \dcirc{0}{-1} \dcirc{0}{-3} \dcirc{0}{-4} \dcirc{0}{-5}
         \dcirc{1}{0} \dcirc{2}{0} \dcirc{4}{0} \dcirc{5}{0} 
         \foreach \i in {-1,-3,-4,-5} {
             \foreach \j in {1,2,4,5} {
                \dcross{\j}{\i} 
            }
        }
         \foreach \i in {0,1,2,4,5} {
             \foreach \j in {0.5,1.5,2} {
                \dcircg{\i+\j}{0+\j} 
             }
         }
         \foreach \i in {0,1,2,4} {
             \foreach \j in {0.5,1.5,2} {
                \dcrossg{5+\j}{-5+\i+\j} 
             }
         }
         \node[anchor=north east] at (0,-5) {\footnotesize$(0,0,0)$};
         \node[anchor=north west] at (5,-5) {\footnotesize$(0,q,0)$};
         \node[anchor=south east] at (0,0) {\footnotesize$(p,0,0)$};
         \node[anchor=south west] at (5+2,0+2) {\footnotesize$(p,q,r+s)$};
         \draw[thick,-{Latex[width=3pt,length=4pt]}] (-0.7,-4.3) -- (-0.7,-4.3+1.6);
         \draw[thick,-{Latex[width=3pt,length=4pt]}] (0.7,-5.7) -- (0.7+1.6,-5.7);
         \draw[gray,thick,-{Latex[width=3pt,length=4pt]}] (5+1.1,-5+0.1+0.2) -- (5+1.1+1.0,-5+0.1+1.0+0.2);
         \node[anchor=east] at (-0.7,-4.3+0.8) {\footnotesize$\oa$};
         \node[anchor=north] at (0.7+0.8,-5.7) {\footnotesize$\ob$};
         \node[anchor=north west] at (5+1.1+0.1,-5+0.1+0.9+0.2) {\rotatebox{45}{\color{gray}\footnotesize$\oc,\od$}};
     \end{tikzpicture}
    \end{center}
    \caption[]{
    Projection of the Hasse diagram with  
          Q-operators on the lattice points $(i,j,k)$. 
          Rational Q-operators are marked with 
          $\begin{tikzpicture}[scale=0.9]\dcirc{0}{0}\end{tikzpicture}$
          and non-rational ones with
          $\begin{tikzpicture}[scale=0.9]\dcross{0}{0}\end{tikzpicture}$ corresponding to truncating and non-truncating $\qlax$-operators respectively.
      }
      \label{fig:distribution}
\end{figure}

The Q-system of $\mathfrak{u}(p,q|r+s)$ contains a total number of
$2^{p+q+r+s}$ Q-operators $\mathbf{Q}_I$ built from the operators $\qlax_I$ with $I\subseteq\{1,\ldots,p+q+r+s\}$. Their functional relations can conveniently be depicted in a Hasse diagram spanned by a hypercube. For our purposes it is convenient to consider a projection of the Hasse diagram onto an ordinary three-dimensional cube. This projection is visualised in Figure~\ref{fig:distribution}. 
In this diagram, each lattice point $(i,j,k)$ is occupied by 
$\binom{p}{i}\times \binom{q}{j}\times \binom{r+s}{k}$ Q-operators, namely those Q-operators whose index set $I$ includes $i$ bosonic indices which are not particle-hole transformed according
to \eqref{eq:realoscs}, $j$ bosonic indices which are transformed, and $k$ fermionic
indices. Here $i=1,\ldots,p$, $j=1,\ldots,q$ and $k=1,\ldots,r+s$.

The total number of indices in the set $I$ is referred to as the level $k=|I|$ which can be thought of as a diagonal slice of the cube. Each level contains $\binom{p+q+r+s}{k}$ Q-operators. Interestingly, for only $2^{r+s}(2^p+2^q-1)$ of them all exponentials in the $\qlax$-operators \eqref{eq:fullax} truncate.
This can be seen from the action of the exponentials on the states defined in \eqref{eq:defstates}.
An $\qlax$-operator $\qlax_I$ has matrix elements with truncating sums if one or both of the following conditions hold:
\begin{itemize}
 \item $I$ does not contain any indices corresponding to bosonic oscillators that are particle-hole transformed.
  \item $I$  contains all indices corresponding to bosonic oscillators that are not particle-hole transformed.
\end{itemize}
Else the matrix elements of the $\qlax$-operator will involve infinite sums.
Due to the nilpotency of the fermionic oscillators, the fermionic degrees of freedom do not change the truncating or non-truncating
nature of the $\qlax$-operators. 
It follows that the truncating $\qlax$-operators are located at the lattice sites
$(i,0,k)$ and $(p,j,k)$ where $i=1,\ldots,p$, $j=1,\ldots,q$ and $k=1,\ldots,r+s$. 
We denote the vertices with truncating ones by  $\begin{tikzpicture}[scale=0.9]\dcirc{0}{0}\end{tikzpicture}$
          and the non-truncating ones by
          $\begin{tikzpicture}[scale=0.9]\dcross{0}{0}\end{tikzpicture}$ in Figure~\ref{fig:distribution}. 
At a given lattice point the latter ones contain $j(p-i)$ pairs of exponentials which do not truncate.

Truncating $\qlax$-operators yield Q-operators whose matrix elements are rational functions of the spectral parameter multiplied by an exponential function including the twist phases. In the case of non-truncating $\qlax$-operators the resulting matrix elements of the Q-operators are written in terms of rational functions and the generalised Lerch transcendent, see Section~\ref{sec:qsys} and Appendix~\ref{sec:formulas}.

\subsection{Ladder decomposition of $\qlax$-operators}

In this section we introduce the decomposition of the $\qlax$-operators \eqref{eq:fullax} on which our approach to evaluate Q-operators for non-compact spin chains is based. Using only algebraic relations we reduce the number of infinite sums and make the actual challenge of evaluating $\qlax$-operators manifest.

Since the factors in the exponentials of the $\qlax$-operators \eqref{eq:fullax}
appear often in the following derivations, we define abbreviations for them:
\begin{equation}
    Y_{\udt a \dt a} = 
        (-1)^{\gr{\udt a}+\gr{\dt a}\gr{\udt a}}
        \oxb_{\udt a \dt a}\ochb_{\udt a}\och_{\dt a}\,,
    \qquad 
    X_{\udt a \dt a} =
    (-1)^{\gr{\dt a}+\gr{\udt a}+\gr{\dt a}\gr{\udt a}}
    \ox_{\dt a \udt a}\ochb_{\dt a}\och_{\udt a}\,.
    \label{eq:defxy}
\end{equation} 
The main idea is to expand the exponentials and to combine terms with
the same difference in the powers of the matching factors 
$X_{\udt a \dt a}$ and $Y_{\udt a \dt a}$ in the exponents.
This can be done using the formula
\begin{equation}
    e^{Y_{\udt a \dt a}}
    f(\NN_{\udt a}, \NN_{\dt a})
    e^{-X_{\udt a \dt a}}
    =
    \sum_{n=-\infty}^{+\infty}
    (Y_{\udt a \dt a})^{\theta(+n)\abs{n}}
    \sum_{k=0}^{\infty}
    \frac{(-1)^k Y_{\udt a \dt a}^k X_{\udt a \dt a}^k}{k! (\abs{n}+k)!}
    f(\NN_{\udt a} - k, \NN_{\dt a} + k)
    (-X_{\udt a \dt a})^{\theta(-n)\abs{n}}
    \label{eq:reordering}
\end{equation}
that can be derived using the oscillator algebra. 
We can furthermore express the factor $Y_{\udt a \dt a}^k X_{\udt a \dt a}^k$ as
\begin{equation}
    Y_{\udt a \dt a}^k X_{\udt a \dt a}^k
    =(-1)^{\gr{\udt a}+\gr{\dt a}}
    \frac{\Gamma(1+\NN_{\udt a \dt a})}{\Gamma(1+\NN_{\udt a \dt a}-k)}
    \frac{\Gamma(1+\NN_{\udt a})}{\Gamma(1+\NN_{\udt a}-k)}
    \frac{\Gamma(\NN_{\dt a}+k+(-1)^{\gr{\dt a}})}
         {\Gamma(\NN_{\dt a}+(-1)^{\gr{\dt a}})}
    \eqndot
    \label{eq:xypowers}         
\end{equation} 
Applying \eqref{eq:reordering} to the $\qlax$-operators
\eqref{eq:fullax}, we find that they can be written in a form which  features a minimal number of creation and annihilation operators,
\begin{equation}
    \qlax_I(z) = 
    \sum_{\{n_{\udt a \dt a}\}=-\infty}^\infty
    \Bigg[
    \prod_{\udt a,\,\dt a}
    (Y_{\udt a\dt a})^{\theta(n_{\udt a \dt a})\abs{n_{\udt a \dt a}}}
    \Bigg]
    \middlepart_I(z , \{\NN\}, \{ n\}  )
    \Bigg[
    \prod_{\udt a,\,\dt a}
    (-X_{\udt a\dt a})^{\theta(-n_{\udt a \dt a})\abs{n_{\udt a \dt a}}}
    \Bigg]
    \eqndot
    \label{eq:multireorder}
\end{equation}
The purely diagonal part $\middlepart_I$ is then given by
\begin{equation}
    \begin{aligned}
        \middlepart_I(z, \{\NN\}, \{ n\}  ) = \sum_{\{k_{\udt a \dt a}\}=0}^\infty
        &\left[ 
            \prod_{\udt a, \dt a}
            \frac{
                (-1)^{(\gr{\udt a}+\gr{\dt a}+1)k_{\udt a \dt a}}
            }{
                \Gamma(k_{\udt a \dt a}+1)
                \Gamma(\abs{n_{\udt a \dt a}}+k_{\udt a \dt a}+1)
            }\frac{\Gamma(1+\NN_{\udt a \dt a})}{ \Gamma(1+\NN_{\udt a \dt a}-k_{\udt a \dt a})}
        \right]
        \\
        &\left[ 
            \prod_{\udt a}
            \frac{
                \Gamma(1+\NN_{\udt a})
            }{
                \Gamma(1+\NN_{\udt a}-\sum_{\dt a}k_{\udt a \dt a})
            }
        \right]
        \left[ 
            \prod_{\dt a}
            \frac{
                \Gamma(
                \NN_{\dt a}+(-1)^{\gr{\dt a}}+\sum_{\udt a}k_{\udt a \dt a}
                )
            }{
                \Gamma(\NN_{\dt a}+(-1)^{\gr{\dt a}})
            }
        \right]
        \\
        & \;\frac{\Gamma(z+1-s_I-\sum_{\dt a}\NN_{\dt a}-\sum_{\udt a, \dt a}k_{\udt a \dt a})}{\Gamma(z+1-s_I-\ccc)}\,.
    \end{aligned}
    \label{eq:multireordermiddle}
\end{equation}

While $\middlepart_I$ looks rather complicated, this representation 
is in fact quite convenient.
First note that for any matrix element of \eqref{eq:multireorder}, 
the outer sums over the variables $n_{\udt a \dt a}$ are always finite, 
and only serve to introduce enough creation and annihilation operators to
produce overlapping states. 
For the lowest and the highest level of the
Q-system, only a single term contributes for any matrix element, which is then effectively given
by the diagonal part.
Compared to naively expressing each term in the exponentials by their power series,
this representation already removes half of the infinite sums.

So far our discussion was purely algebraic and we did not specify the spectrum of the number operators.
Assuming that these operators act on a Hilbert space as given
in \eqref{eq:defstates}, the spectrum of the $\NN_{\udt a}$ is positive or negative integer valued depending on whether the corresponding oscillators are particle-hole transformed, cf. \eqref{eq:realoscs}.
For compact representations, all $\NN$ are positive or zero and the sums over the variables $k_{\udt a \dt a}$ in \eqref{eq:multireordermiddle} are finite; they are effectively truncated by the Gamma functions in the
denominator. However, for non-compact representations some $\NN$ take negative integer values
such that some of the sums may not truncate.
The evaluation of those sums is however simplified by the fact that they only involve
diagonal operators.

\subsection{Integral representation for lowest level $\qlax$-operators}
\label{sec:lowest}
In this subsection we focus on the $\qlax$-operators of the lowest level $\qlax_{\{\udt a\}}$. 
For these $\qlax$-operators only one term in the sum \eqref{eq:multireorder} contributes
to any matrix element, which is then directly given by the diagonal part $\middlepart_{\{a\}}$.
Furthermore, as will be discussed in Section~\ref{sec:qsys}, the corresponding Q-operators determine the full Q-system. Here we derive an integral representation of the
diagonal part \eqref{eq:multireordermiddle}, which can easily be evaluated
and from which rational finite sum expressions can readily be obtained, cf.~Section~\ref{sec:derivation} and \ref{sec:schml}.

We first specialise the expression given in \eqref{eq:multireorder} and 
\eqref{eq:multireordermiddle} to the lowest level, and 
write the Lax operators as
\begin{equation}
    \qlax_{\{\udt a\}}(z) =
    \sum_{\{n_{\dt a}\}=-\infty}^\infty
    \left[
        \prod_{\dt a} 
        (Y_{\udt a \dt a})^{\theta(+n_{\dt a})\abs{n_{\dt a}}}
    \right]
    \middlepart_{\{\udt a\}}(z,\{\NN\}, \{ n\})
    \left[
        \prod_{\dt a} 
        (-X_{\udt a \dt a})^{\theta(-n_{\dt a})\abs{n_{\dt a}}}
    \right]
    \eqncom
    \label{eq:form_lax_lowest}
\end{equation}
with $X$ and $Y$ given in \eqref{eq:defxy}. Here the diagonal part reads
\begin{equation}
    \begin{aligned}
    \middlepart_{\{\udt a\}}(z,\{\NN\}, \{ n\})
    =
    \sum_{\{k_{\dt a}\}=0}^{\infty}
    &
    \prod_{\dt a} 
    \Bigg[
        \frac{
            (-1)^{(\gr{\udt a}+ \gr{\dt a}+1)k_{\dt a}}
            \Gamma(1+\NN_{\udt a \dt a})
        }{
            k_{\dt a}! \, (\abs{n_{\dt a}}+k_{\dt a})! \;
            \Gamma(1+\NN_{\udt a \dt a}-k_{\dt a})
        }
        \frac{
            \Gamma(\NN_{\dt a}+(-1)^{\gr{\dt a}}+k_{\dt a})
        }{
            \Gamma(\NN_{\dt a}+(-1)^{\gr{\dt a}})
        }
    \Bigg]
    \\ &\times
    \frac{
        \Gamma(\NN_{\udt a}+1)
    }{
        \Gamma(\NN_{\udt a}+1-\textstyle\sum\nolimits_{\dt a}k_{\dt a})
    }
    \frac{
        \Gamma(z+1-\textstyle\sum\nolimits_{\dt a}(\NN_{\dt a}+k_{\dt a}-\sfrac{1}{2}(-1)^{\gr{\dt a}}))
    }{
        \Gamma(z+1-\ccc-\textstyle\frac{1}{2}\sum_{\dt a}(-1)^{\gr{\dt a}})
    }
    \eqndot
    \end{aligned}
    \label{eq:middlepartlowest}
\end{equation}
To obtain the aforementioned integral representation, we evaluate all sums over the variables 
$k_{\dt a}$ in the diagonal part \eqref{eq:middlepartlowest}.
Since the intermediate formulas are quite lengthy, we only sketch this derivation.

Consider the first sum, which we take to be over some index $\dt b$.
It is straightforward to see that this sum can be written as a product
involving Gamma functions and the following hypergeometric function:
\begin{equation}
    \pFq[36]{3}{2}{%
        \sum\nolimits_{\dt a \neq \dt b}k_{\dt a}-\NN_{\udt a},%
        -\NN_{\udt a \dt b},%
        \NN_{\dt b}+(-1)^{\gr{\dt b}}%
    }{%
        -z +{\textstyle\frac{1}{2}\sum_{\dt a}(-1)^{\gr{\dt a}}}  + \sum\nolimits_{\dt a \neq \dt b}(\NN_{\dt a}+k_{\dt a})+\NN_{\dt b},%
        1+\abs{n_{\dt b}}%
    }{(-1)^{\gr{\udt a}+\gr{\dt b}}}
    \eqndot
    \label{eq:hyper}
\end{equation}
Since the other summation variables appear in the arguments of this hypergeometric function,
the other sums cannot be taken easily. To remedy this, and to disentangle
the sums, one can use an integral representation of the hypergeometric function.
The type of integral however depends on the spectrum of the operator $\NN_{\udt a}$.
If the oscillator with index $\udt a$ is bosonic and particle-hole transformed,
$\omega_{\udt a}=-1$, the first argument $\sum_{\dt a \neq \dt b}k_{\dt a}-\NN_{\udt a}$
of the hypergeometric function \eqref{eq:hyper} takes positive integer values, and we can use 
the standard Euler type integral, expressing the function ${}_3F_2$ as 
an integral over the interval
$(0,1)$ on the real line involving the function ${}_2F_1$.
For all other cases, the Gamma function 
$\Gamma(\NN_{\udt a}+1-\textstyle\sum\nolimits_{\dt a}k_{\dt a})$ in the
denominator of \eqref{eq:middlepartlowest} truncates the range of the
summation variables such that the argument 
$\sum_{\dt a \neq \dt b}k_{\dt a}-\NN_{\udt a}$ of the hypergeometric function \eqref{eq:hyper}
takes non-positive integer values. For negative arguments, one can use an analytic
continuation of the Euler integral employing the Pochhammer contour. For negative
integers, this contour collapses into a contour integral around the origin.

Using the appropriate integral formulas to rewrite the hypergeometric function \eqref{eq:hyper},
one finds that all subsequent summations decouple, and can be performed easily
using the series representation of the hypergeometric function ${}_2F_1$.
We then arrive at the result
\begin{equation}
    \begin{aligned}
        &
    \middlepart_{\{\udt a\}}(z,\{\NN\}, \{ n\}) \\&= 
    \int\dd t \;
    t^{-\NN_{\udt a }-1} (1-t)^{-z-1+\ccc +\frac{1}{2}\sum_{\dt a}(-1)^{\gr{\dt a}}}
    \prod_{\dt a} \frac{1}{\abs{n_{\dt a}}!}\;
    \pFq{2}{1}{\NN_{\dt a} + (-1)^{\gr{\dt a}}, -\NN_{\udt a \dt a}}{1+\abs{n_{\dt a}}}{(-1)^{\gr{\dt a}+\gr{\udt a}}t}
    \eqncom
\end{aligned}
    \label{eq:middlepart_lowest}
\end{equation}
where $\ccc$ is the central charge defined in \eqref{eq:def_central_charge} and the integration is
\begin{equation}
    \int \dd t = \begin{cases}\displaystyle
        \frac{(-1)^{\NN_{\udt a}} }{\Gamma(-\NN_{\udt a})} \int_0^1 \dd t 
        & \quad \text{if } \gr{\udt a}=0 \text{ and } \omega_{\udt a} = -1 \\[20pt]
        \displaystyle
        \frac{\Gamma(1+\NN_{\udt a})}{2\pi i} \oint_{t=0} \dd t
        & \quad \omega_a=1\text{ or }\gr{a}=1
    \end{cases}
    \eqndot
    \label{eq:def_int_lowest_new}
\end{equation}
This means that for truncating $\qlax$-operators, the integral just computes
a residue, while for the non-truncating ones, it is an integral over
the interval $(0,1)$.
Note that strictly speaking, the integral is only convergent for appropriately
chosen values of the spectral parameter $z$; this however poses no problem,
since the result for any matrix element is a rational function which can
be analytically continued to any value of the spectral parameter.

Further we note that while it might seem that we did not gain much by writing the potentially 
infinite sums of the $\qlax$-operator in terms of an integral, this
integral is in fact trivial to evaluate, by expanding the integrand and, depending on the case,
either taking a simple residue or evaluating the line integral in terms
of a Beta function. It provides a convenient way of treating the truncating and 
non-truncating $\qlax$-operators in a joint way, the essentially only difference
being the contour of integration.
In the next section we show how the integral formula \eqref{eq:middlepart_lowest} is used to recover the matrix elements in the case of the spin~$-s$ chains discussed in Section~\ref{sec:sl2}. Subsequently, we demonstrate how the integral formula can be rewritten in terms of finite sums in Section~\ref{sec:schml}.

\subsection{Derivation of the $\qlax$-operators for the spin $-s$ models}
\label{sec:derivation}

The integral representation for $\qlax$-operators of the lowest level,
given in \eqref{eq:form_lax_lowest} 
together with \eqref{eq:middlepart_lowest}, can easily be
evaluated in practise. 
To show that it also serves as a good starting point to obtain representations in terms of finite sums, we now derive the formulas
\eqref{eq:formR2}, 
\eqref{eq:middleR2},
\eqref{eq:formR1} and
\eqref{eq:middleR1}
for the $\qlax$-operators of the spin~$-s$ spin 
chains considered in Section~\ref{sec:sl2}.
For these models, both oscillators are bosonic,
$\gr{\cdot}=(0,0)$, and the first oscillator is particle-hole transformed,
$\omega=(-1,+1)$. The central charge is constrained to $\ccc=-2s$, such that
the states are given by $\ket{m}_s=\ket{2s-1+m,m}$, cf. \eqref{eq:statesrank1}.

We begin with the truncating  $\qlax$-operator $\qlax_{\{2\}}$.
The matrix elements ${}_s\bra{\tilde m}\qlax_{\{2\}}(z)\ket{m}_s$
can be determined from  \eqref{eq:form_lax_lowest} 
by noting that the summation variable $n_1$ is fixed to be $n_1=\tilde m -m$,
and that the diagonal part then acts on a state $\ket{\mmz}_s$ with $\mmz=\min(m,\mt)$, cf.~\eqref{eq:fixns_one} and \eqref{eq:m0s_one}.
Using this it is straightforward to show that the general structure of the
Lax operator exactly matches \eqref{eq:formR2}.
The diagonal part $\middlepart_{\{2\}}$ could in principle be derived directly 
from expression \eqref{eq:middlepartlowest}; we nevertheless start from the 
generally applicable formula \eqref{eq:middlepart_lowest} expressing
it as a contour integral. This integral can be evaluated by plugging in 
the series representations of the hypergeometric function and of the power of
$(1-t)$,
\begin{align}
    \allowdisplaybreaks[1]
    &{}_s\bra{\mmz}\middlepart_{\{2\}}(z)\ket{\mmz}_s
    \nonumber\\[6pt]
    &=
    \frac{\mmz!}{\abs{m-\mt}!}
    \frac{1}{2\pi i} \oint_{t=0}\dd t \;
    t^{-\mmz-1}(1-t)^{-z-\frac{1}{2}-2s}
    \pFq{2}{1}{1-2s-\mmz, -\NN_{21}}{1+\abs{\mt-m}}{t}
    \nonumber\\[6pt]
    &=
    \mmz!
    \frac{1}{2\pi i} \oint_{t=0}\dd t \;
    t^{-\mmz-1}
    \left[
        \sum_{\ell=0}^\infty \frac{(z+\frac{1}{2}+2s)_\ell}{\ell!}t^\ell
    \right]
    \left[
        \sum_{k=0}^{\mmz+2s-1} \frac{
            (2s+\mmz-k)_k(1+\NN_{21}-k)_k
        }{
            (\abs{\mt-m}+k)!
            k!
        }
        t^k
    \right]
    \nonumber\\[6pt]
    &=
    \sum_{k=0}^{\mmz} 
    \binom{\mmz}{k}
    \frac{
        (z+\frac{1}{2}+2s)_{\mmz-k}
        (1+\NN_{21}-k)_k
        (2s+\mmz-k)_k
    }{
        (\abs{\mt-m}+k)!
    }
    \eqncom
\end{align}
which is the same as \eqref{eq:middleR2}.

Next we turn to the non-truncating $\qlax$-operator $\qlax_{\{1\}}$.
For each matrix element, the summation variable is fixed to $n_2=m-\mt$,
and the diagonal part acts on $\ket{\mmz}_s$, where now $\mmz=\max(m,\mt)$.
One finds that the form of the matrix elements in \eqref{eq:formR1} is reproduced by 
\eqref{eq:form_lax_lowest}.
The diagonal part 
\eqref{eq:middlepart_lowest}
is then given by
\begin{equation}
    \begin{aligned}
        {}_s\bra{\mmz}\middlepart_{\{1\}}(z)\ket{\mmz}_s&=
    \frac{(-1)^{2s+\mmz}}{(2s-1+\mmz)!\abs{\mt-m}!}\int_0^1 \dd t \;
    t^{2s-1+\mmz}(1-t)^{-z-\frac{1}{2}-2s}
    \pFq{2}{1}{\mmz+1, -\NN_{12}}{1+\abs{\mt-m}}{t}\,.
    \end{aligned}
\end{equation}
To write the matrix elements as finite sums, we have to apply the Euler
transformation 
$ \,_2F_1(n,b;m;z)=(1-z)^{m-n-b} {}_2F_1(m-n,m-b;m;z) $
to the hypergeometric function. Then this function can
be written as a finite sum and the integral can be evaluated using the
integral representation of the Beta function:
\begin{align}
    \allowdisplaybreaks[1]
    %&
    {}_s\bra{\mmz}\middlepart_{\{1\}}(z)\ket{\mmz}_s
    %\nonumber\\[6pt]
    &=
    \frac{(-1)^{2s+\mmz}}{(2s-1+\mmz)!\abs{\mt-m}!}
    \int_0^1 \dd t \;
    t^{2s-1+\mmz}(1-t)^{-z-\frac{1}{2}-2s-\min(m,\mt)+\NN_{12}}
    \nonumber\\&\qquad\qquad\qquad\qquad\times
    \pFq{2}{1}{-\min(m\text{,}\,\mt), 1+\abs{\mt-m}+\NN_{12}}{1+\abs{\mt-m}}{t}
    \nonumber\\[6pt]
    &=
    \sum_{k=0}^{\min(m,\mt)}
    \frac{(-1)^{2s+\mmz}}{(2s-1+\mmz)!}
    \frac{
    (1+\min(m,\mt)-k)_k
    (1+\abs{\mt-m}+\NN_{12})_k
    }{
        k! (\abs{\mt-m}+k)!
    }
    \nonumber\\&\qquad\qquad\qquad\qquad\times
    \int_0^1 \dd t \;
    t^{2s-1+\mmz+k}(1-t)^{-z-\frac{1}{2}-2s-\min(m,\mt)+\NN_{12}}
    \nonumber\\[6pt]
    &=
    \sum_{k=0}^{\min(m,\mt)}
    \frac{(-1)^{2s+\mmz}}{(2s-1+\mmz)!}
    \frac{
    (1+\min(m,\mt)-k)_k
    (1+\abs{\mt-m}+\NN_{12})_k
    }{
        k! (\abs{\mt-m}+k)!
    }
    \nonumber\\&\qquad\qquad\qquad\qquad\times
    B(2s+\mmz+k,-z+\sfrac{1}{2}-2s-\min(m,\mt)+\NN_{12})
    \eqndot
\end{align}
This expression is identical to \eqref{eq:middleR1}, upon using $B(x,y)=\frac{\Gamma(x)\Gamma(y)}{\Gamma(x+y)}$.

In the next section we represent the generalisation of this finite sum formula
for the non-truncating $\qlax$-operators of arbitrary non-compact spin chains of Jordan-Schwinger type.

\subsection{Finite sum representation for lowest level $\qlax$-operators}\label{sec:schml}

Evaluating the integral formula in \eqref{eq:middlepart_lowest} 
is a very efficient way to determine matrix elements of truncating as well as
non-truncating $\qlax$-operators. 
It is however also possible to directly derive finite sum expressions from the
integral representation using the same ideas as in the previous section.
Here one has to treat the truncating and non-truncating $\qlax$-operators separately,
corresponding to the two integration contours in \eqref{eq:def_int_lowest_new}.
In the truncating case, evaluating the residue returns the expression given in
\eqref{eq:middlepartlowest}, which can be expressed in terms of number operators
for the particle-hole transformed oscillators \eqref{eq:realoscs}; then all sums
are manifestly finite.

For the non-truncating $\qlax$-operators $\qlax_{\{a\}}$ with $\gr{\udt a}=0$ and 
$\omega_a=-1$, one can evaluate the integral as follows: 
First, we decompose the set into sets of indices corresponding to the different types of oscillators $\bar I = 
\bar I_\oa \cup
\bar I_\ob \cup
\bar I_\oc \cup
\bar I_\od 
$, cf.~\eqref{eq:realoscs}.
Subsequently we apply the Euler transformation to the hypergeometric functions corresponding to 
the set $\bar I_{\oa}$ and use the series expansion of all such functions to perform the Beta integral. We find
\begin{equation}
    \begin{aligned}
        &\middlepart_{\{\udt a\}}(z,  \{\NN\}, \{ n\}  )  \\[5pt]
    &= \sum_{\{k_{\dt a}\}=0}^\infty 
    \frac{(-1)^{1+\NN_{\ob_{\udt a}}}}{\NN_{\ob_{\udt a}}!}
    \frac{1}{
        \prod_{\dt a\in\bar I } k_{\dt a}! (\abs{n_{\dt a}}+k_{\dt a})!
    }
    \\
    &\qquad\qquad
    \prod_{\dt a \in \bar I_{\oa}}
    (\abs{n_{\dt a}}-\NN_{\oa_{\dt a}})_{k_{\dt a}}
    (\abs{n_{\dt a}}+\NN_{\udt a \dt a}+1)_{k_{\dt a}}
    \qquad
    \prod_{\dt a \in \bar I_{\ob}}
    (-\NN_{\ob_{\dt a}})_{k_{\dt a}}
    (-\NN_{\udt a \dt a})_{k_{\dt a}}
    \\&\qquad\qquad
    \prod_{\dt a \in \bar I_{\oc}}
    (-1)^{k_{\dt a}}
    (\NN_{\oc_{\dt a}}-1)_{k_{\dt a}}
    (-\NN_{\udt a \dt a})_{k_{\dt a}}
    \qquad
    \prod_{\dt a \in \bar I_{\od}}
    (-1)^{k_{\dt a}}
    (-\NN_{\od_{\dt a}})_{k_{\dt a}}
    (-\NN_{\udt a \dt a})_{k_{\dt a}}
    \\[3pt]
        & \qquad\qquad
        B\Big(
        -z+\ccc+\sfrac{1}{2}\sum_{\dt a \in \bar I}(-1)^{\gr{\dt a}}+\sum_{\dt a \in \bar I_{\oa}}(\NN_{\udt a\dt a}+\abs{n_{\dt a}}-\NN_{\oa_{\dt a}})
        ,
        \NN_{\ob_{\udt a}}+1+\sum_{\dt a \in \bar I} k_{\dt a}
        \Big)
        \eqncom
\end{aligned}
\label{eq:finite_sum_nonpolynomial}
\end{equation}
where we denoted the number operators of the particle-hole 
transformed oscillators as $\NN_{\oa_{a}}=\oab_a\oa_a$ et cetera.
All the Pochhammer symbols involving these operators are of the form 
$(-m)_{k}$ with $m\geq 0$ which gives $(-m)_k=(-1)^k\frac{\Gamma(m+1)}{\Gamma(m+1-k)}$,
such that all sums truncate. Note that the fact that
$\abs{n_{\dt a}-\NN_{\oa_{\dt a}}} \leq 0$
can be seen from the structure of the outer sums in \eqref{eq:form_lax_lowest},
see also Appendix~\ref{sec:matrixelements}.

\subsection{Towards higher level finite sum representations}
\label{sec:higher}

We have seen that the $\qlax$-operators of the lowest level
can conveniently be written either using the integral representation \eqref{eq:middlepart_lowest}
or as finite sums as in \eqref{eq:middlepartlowest} and \eqref{eq:finite_sum_nonpolynomial}.
Here we want to discuss the generalisation of such representation to the remaining 
levels of the Q-system.

First note that the $\qlax$-operators are almost symmetric under the
exchange $I\leftrightarrow\bar I$; it is therefore possible to proceed with the highest level $\qlax$-operators similarly as with the lowest level ones. These results are summarised in Appendix~\ref{sec:highest}.
The intermediate levels can be much more involved. However, we note that the difficulty of deriving representations
without infinite sums does not necessarily increase according to the level $|I|$ of the
operators, but rather by the number of infinite sums, or more precisely by
the number of indices $a\in I$ which correspond to bosonic and particle-hole
transformed oscillators, $\gr{a}=0$ and $\omega_a=-1$.
If no such indices appear in the index set of $\qlax_I$, the formula \eqref{eq:multireordermiddle}
contains no infinite sum to begin with, cf.~Section~\ref{sec:com}.
Furthermore, for the case that there is exactly one such index, 
one can apply the same strategy as was used for the lowest level.
Let this index be $\udt b$; then one can perform all sums 
over the variables $k_{\udt b \dt a}$ in \eqref{eq:multireordermiddle},
and obtain a formula with finite sums and an integral as in Section~\ref{sec:lowest}.
Writing the integral in terms of finite sums as in Section~\ref{sec:schml}, one obtains
a formula in terms of finite sums only.

The first case where more severe difficulties arise can best be discussed
using a concrete example.
Consider a $\mathfrak{u}(2,2)$ invariant model, with oscillators
$\oa_1$,
$\oa_2$,
$\ob_3$ and
$\ob_4$.
Then the operator $\qlax_{\{3,4\}}$ contains two particle-hole transformed indices.
After performing similar calculations as for the lowest level, one 
finds the following representation:
\begin{equation}
\begin{split}
    \qlax_{\{3,4\}}(z)&=e^{-\sum_{a,\bar a}\oxb_{a\bar a} \oa_a \ob_{\bar a}}
    \;
    \frac{
    \Gamma(z-\NN_{\oa_1}-\NN_{\oa_2})
    }{
        \Gamma(z-\ccc)
    }\;
    e^{-\sum_{a, \bar a}\ox_{\bar a a} \oab_a\obb_{\bar a}}\\ &=\sum_{\{n_{a\bar a }\}=-\infty}^{+\infty}\left[\prod_{a,\bar a} \left(-\oxb_{a\bar a}\oa_a\ob_{\bar a}\right)^{\theta(+n_{a\bar a})|n_{a\bar a}|}\right]
    \middlepart_{\{3,4\}}(z , \{\NN\}, \{ n\})  )
    \left[\prod_{a,\bar a} \left(-\ox_{\bar a a}\oab_a\obb_{\bar a}\right)^{\theta(-n_{a\bar a})|n_{a\bar a}|}\right]\,,
 \end{split}
\end{equation} 
where the indices run over $\udt a\in I=\{1,2\}$ and $\dt a\in \bar I=\{3,4\}$.
Here the diagonal part $\middlepart_{\{3,4\}}(z)$ can be written in terms of finite sums and an integral,
\begin{equation}
\small
\begin{split}
&\middlepart_{\{3,4\}}(z, \{\NN\}, \{ n\})
 = 
 \sum_{k_{13},k_{23}=0}^\infty \;
 \frac{
     (-1)^{\NN_{\ob_3}+\NN_{\ob_4}}
  }{
      n_{14}!n_{24}!
      \NN_{\ob_3}!\NN_{\ob_4}!
  }
  \;
  \frac{
     (k_{13}+k_{23}+\NN_{\ob_3})!
  }{
     k_{13}!k_{23}!(k_{13}+n_{13})!(k_{23}+n_{23})!
  }
  \\[8pt] & \times
 \frac{
     (n_{13}-\NN_{\oa_1})_{k_{13}}
     (n_{23}-\NN_{\oa_2})_{k_{23}}
     (1+n_{13}+\NN_{13})_{k_{13}}
     (1+n_{23}+\NN_{23})_{k_{23}}
 }{
     (n_{13}+n_{23}+\NN_{13}+\NN_{23}-\NN_{\ob_3}+1-z)_{k_{13}+k_{23}+\NN_{\ob_3}}
 }
 \\[3pt] 
&\times
\int_0^1t^{\NN_{\ob_4}}(1-t)^{\ccc -z}
 \;\; \,_3F_2(\NN_{\oa_1}+1,1-n_{13}+\NN_{\oa_1},-\NN_{14};1-k_{13}-n_{13}+\NN_{\oa_1},1+n_{14};t)
 \\
 &\;\;\,\qquad\qquad\qquad\qquad\times\,_3F_2(\NN_{\oa_2}+1,1-n_{23}+\NN_{\oa_2},-\NN_{24};1-k_{23}-n_{23}+\NN_{\oa_2},1+n_{24};t)
 \end{split}
\end{equation} 
where the central charge is $\ccc=\NN_{\oa_1}+\NN_{\oa_2}-\NN_{\ob_3}-\NN_{\ob_4}-2$.
It resembles the integral formula \eqref{eq:middlepart_lowest}, but this time involving generalised hypergeometric functions. 
Indeed it is even possible to write the integral in terms of finite sums, using 
an analogue of the Euler transformation which can be found in \cite{miller2013}. 
This identity is however rather involved and not very explicit, and requires
finding the zeros of an auxiliary polynomial. 
Note that the formula for $\middlepart_{\{3,4\}}(z)$ follows from 
first using the result \eqref{eq:finite_sum_nonpolynomial} 
for finite sum representations of the lowest levels to make half of the
sums finite. Then one applies the same strategy to the remaining sums.
The fact that the next step requires implicit identities of the type just discussed for the
hypergeometric functions renders it difficult to treat cases with more infinite sums 
in this recursive fashion.

Nevertheless, as discussed in the next section, for the purpose of calculating
Q-operators explicitly there is no need
to evaluate higher level $\qlax$-operators as the whole Q-system can be obtained from the lowest level.

%%%%%%%%%%%%%%%%%%%%%%%%%%%%%%%%%%%%%%%%%%%%%%%%%%%%%%%%%%%%%%%%%%%%%%%%%%%%%%%
%%%%%%%%%%%%%%%%%%%%%%%%%%%%%%%%%%%%%%%%%%%%%%%%%%%%%%%%%%%%%%%%%%%%%%%%%%%%%%%
%%%%%%%%%%%%%%%%%%%%%%%%%%%%%%%%%%%%%%%%%%%%%%%%%%%%%%%%%%%%%%%%%%%%%%%%%%%%%%%
\section{Generating the operatorial Q-system}
\label{sec:qsys}

Above we focused on the calculation of matrix elements for  lowest level $\qlax$-operators. In fact, as we will discuss now, the lowest level $\qlax$-operators are sufficient to generate the entire operatorial
Q-system. 
Our strategy is to first combine the $\qlax$-operator's matrix elements into matrix
elements of the respective Q-operator by taking products and tracing
out the auxiliary Fock space.
For each magnon block, the Q-operators can be represented explicitly as
matrices of finite size.
Systematically solving the functional relations \eqref{eq:QQb} and \eqref{eq:QQf}, we determine all other 
Q-operators in the corresponding magnon block.
To facilitate concrete calculations, we collect all formulas 
necessary in this process in Appendix~\ref{sec:formulas}.
This allows to perform all calculations in computer algebra systems such as Mathematica.

\subsection{Lowest level Q-operators}

Using the matrix elements of the $\qlax$-operators of the lowest level 
one can construct matrix elements of the corresponding Q-operators via \eqref{eq:qop}. Due to the remaining $\mathfrak{u}(1)^{N+M}$ invariance which persists in the 
presence of diagonal twists, 
the Q-operators are block diagonal. 
These blocks correspond to sectors with a fixed number of magnons;
they are labelled by the total excitation numbers $\sum_{i=1}^L m_a^{(i)}$, where
$a=1,\ldots,p+q+r+s$,
of the oscillators of the representation of $\mathfrak{u}(p,q|r+s)$ given in \eqref{eq:realoscs},
see also \eqref{eq:defstates},
and the number of sites $L$.
For each such magnon block, 
the matrix elements can therefore be combined into a matrix of finite size.
This gives the operatorial form of Q-operators in a subspace of the 
infinite-dimensional Hilbert space of non-compact models. 
For spin chains of length $L$ the matrix elements of the lowest level Q-operators can be expressed as
\begin{multline}
    \Big(
    \bra{\mtb^{(L)}}
    \cdots
    \bra{\mtb^{(1)}}
    \Big)
    \;
    \qop_{\{a\}}(z)
    \;
    \Big(
    \ket{\mb^{(1)}}
    \cdots
    \ket{\mb^{(L)}}
    \Big)
    \\[8pt]
    =(-1)^{\sum_{i<j}\gr{\mtb^{(j)}}(\gr{\mb^{(i)}}+\gr{\mtb^{(i)}})}\;
    e^{iz\phi_{a}}\;
    \widehat \str\,
    \bra{\mtb^{(1)}}\qlax_{\{a\}}(z)\ket{\mb^{(1)}}
    \cdots
    \bra{\mtb^{(L)}}\qlax_{\{a\}}(z)\ket{\mb^{(L)}}\,.
    \label{eq:meQ}
\end{multline} 
Here we denote the Graßmann degree of the state $\ket{\mb^{(i)}}$ defined in \eqref{eq:defstates} by $\gr{\mb^{(i)}}$.
The matrix elements of the $\qlax$-operators     $\bra{\mtb^{(i)}}\qlax_{\{a\}}(z)\ket{\mb^{(i)}}$ follow immediately from the
integral representation given in \eqref{eq:form_lax_lowest} and \eqref{eq:middlepart_lowest},
or the finite sum for the non-truncating $\qlax$-operators given in \eqref{eq:finite_sum_nonpolynomial}.
They can be found in full detail
in Appendix~\ref{sec:matrixelements}.
Of course, these matrix elements still depend on the auxiliary space operators
$\oxb_{\udt a \dt a}$ and $\ox_{\dt a \udt a}$.

To evaluate \eqref{eq:meQ}, one first commutes all the auxiliary space operators
either to the left or to the right, and combines
them into number operators $\NN_{\udt a \dt a}$.
All terms containing any off-diagonal terms, i.e. raising or lowering operators,
can then be dropped since they do not contribute to the supertrace.
The normalised supertrace is then given in terms of ordinary 
sums over these remaining diagonal terms, which however need to be regularised,
by giving the twist angles small imaginary parts, as discussed in Section~\ref{sec:qs}.
Note that the definition of the trace \eqref{eq:str} factors into traces over the 
individual Fock spaces of the different auxiliary 
space oscillators, $\widehat\str=\prod_{a,b}\widehat\str_{ab}$,
where $\widehat\str_{ab}$ traces out the oscillator $(\oxb_{ab},\ox_{ba})$.
One finds that only a closed set of a few different types of sums can occur
when calculating the traces $\widehat\str_{ab}$,
including sums over rational functions and the Lerch transcendent \eqref{eq:def_hl}.
Formulas for all these sums are collected in Appendix~\ref{sec:str}.

\subsection{Operatorial Q-system from functional relations}
\label{sec:abcde}

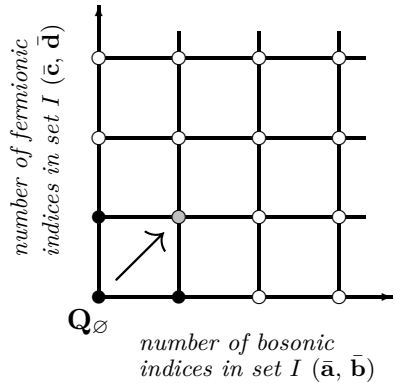
\begin{figure}[t]
\centering
\begin{picture}(120,130)

    \put(-12,-12){\itshape $\qop_\es$}

    \put(15,-20){\footnotesize\itshape number of bosonic}
    \put(15,-30){\footnotesize\itshape indices in set $I$ \normalfont($\oab$, $\obb$)}

    \put(-33,15){\rotatebox{90}{\footnotesize\itshape number of fermionic}}
    \put(-23,15){\rotatebox{90}{\footnotesize\itshape indices in set $I$ \normalfont($\ocb$, $\odb$)}}

\linethickness{0.4mm}
\put(0,0){\vector(1,0){110}}
\put(0,30){\line(1,0){100}}
\put(0,60){\line(1,0){100}}
\put(0,90){\line(1,0){100}}

\put(0,0){\vector(0,1){110}}
\put(30,0){\line(0,1){100}}
\put(60,0){\line(0,1){100}}
\put(90,0){\line(0,1){100}}

\put(0,0){\circle*{5}}
\put(30,0){\circle*{5}}
\put(0,30){\circle*{5}}

\put(60,0){\color{white}\circle*{5}}\put(60,0){\circle{5}}
\put(90,0){\color{white}\circle*{5}}\put(90,0){\circle{5}}

\put(30,30){\color{white}\circle*{5}}\put(30,30){\color{lightgray}\circle*{5}}\put(30,30){\circle{5}}
\put(60,30){\color{white}\circle*{5}}\put(60,30){\circle{5}}
\put(90,30){\color{white}\circle*{5}}\put(90,30){\circle{5}}

\put(0,60){\color{white}\circle*{5}}\put(0,60){\circle{5}}
\put(30,60){\color{white}\circle*{5}}\put(30,60){\circle{5}}
\put(60,60){\color{white}\circle*{5}}\put(60,60){\circle{5}}
\put(90,60){\color{white}\circle*{5}}\put(90,60){\circle{5}}

\put(0,90){\color{white}\circle*{5}}\put(0,90){\circle{5}}
\put(30,90){\color{white}\circle*{5}}\put(30,90){\circle{5}}
\put(60,90){\color{white}\circle*{5}}\put(60,90){\circle{5}}
\put(90,90){\color{white}\circle*{5}}\put(90,90){\circle{5}}

\linethickness{1mm}
\put(5,10){\color{black}\huge$\nearrow$}

\end{picture}
\vspace{2.5\baselineskip}
\caption{Generation of the full Q-system from $\qop_{\es}$ and the set of $\qop_{\{a\}}$ at black nodes. 
The arrow signals the need to solve the difference equation \eqref{eq:guemmel} to obtain the Q-operators on the grey node.
All Q-operators on the white nodes can then be obtained from the determinant formulas \eqref{eq:detformulas}. 
The lattice shown here is a projection of the one used in Figure~\ref{fig:distribution}.}
\label{fig:solveQQ}
\end{figure}

Knowing the Q-operators with a single index as explicit matrices for a 
given magnon block, one can produce explicit matrices for all operators of higher level
by imposing the QQ-relations \eqref{eq:QQb} and \eqref{eq:QQf}.
\footnote{
     Using the approach we 
    present here, recovering the known expression for $\qop_{\fs}$ given in
    \eqref{eq:qemptyfull} constitutes a non-trivial check of these relations, which we performed 
    for specific examples.
}
A naive way of solving the bosonic relation \eqref{eq:QQb} however involves
a matrix inversion, which is problematic given that the Q-operators 
are expressed in terms of special functions.

A more efficient strategy is to first calculate the Q-operators with 
one bosonic and one fermionic index, $\qop_{\{a,b\}}$ with $|a|\neq |b|$.
To obtain these, we need to solve the first order difference equation given by \eqref{eq:QQf}:
\be
\qop_{\{a,b\}}(z)-\qop_{\{a,b\}}(z+1)=-\Delta_{ab}\qop_{\{a\}}(z+\sfrac{1}{2})\qop_{\{b\}}(z+\sfrac{1}{2})\quad\quad |a|\neq |b|\,.
\label{eq:guemmel}
\ee
The formal solution to this equation can be written in terms of the discrete 
analogue of integration, which we denote by $\Sigma$ and define through $\Sigma\left[ f(z)-f(z+1) \right]=f(z)+\mathcal{P}$. Here $\mathcal{P}$ is periodic, $\mathcal{P}(z)=\mathcal{P}(z+1)$.
The discrete integral can be written as a sum, $\Sigma[f(z)]=\sum_{n=0}^\infty f(z+n)$,
whenever this sum converges.
For the Q-operators with one bosonic and one fermionic index we can thus write
\be
\qop_{\{a,b\}}(z) = -\Delta_{ab}\;\Sigma\left[ \qop_{\{a\}}(z+\sfrac{1}{2})\qop_{\{b\}}(z+\sfrac{1}{2}) \right]\quad\quad |a|\neq |b|\,.
\label{eq:finite_diff}
\ee
We describe the explicit realisation of this operation on the encountered basis of functions in  Appendix~\ref{sec:Psi}, where we also make it clear that all Q-operators are given in terms of linear combinations of rational functions and generalised Lerch transcendents which are likewise defined there.
It is important to note that in contrast to the untwisted case, 
the arbitrary periodic function $\mathcal{P}$ is fixed to be zero if we require
the Q-operators obtained from \eqref{eq:finite_diff} to be identical 
to the monodromy construction, since $\mathcal{P}$ is incompatible with the
exponential scaling in terms of the twist phases.

Via the QQ-relations, it is possible to write all other Q-operators
as determinants of $\qop_{\{a\}}$ and $\qop_{\{a,b\}}$ with $|a|\neq|b|$,
\be
&&\qop_{\{a_1,\ldots,a_m,b_1,\ldots,b_n\}} =
\frac{\prod_{i=1}^m\prod_{j=1}^n\Delta_{a_i b_j}}{\prod_{1\le i<j\le m} \Delta_{a_i a_j} \prod_{1\le i<j\le n} \Delta_{b_i b_j}  } \\[10pt]
&&\quad\quad \times \left\{
\begin{matrix}
    (-1)^{(n-m)m}
\epsilon^{k_1,...,k_n}
 \prod_{r=1}^{m} 
\frac{1}{
 \Delta_{a_r b_{k_r} }
}
\qop_{\{a_r, b_{k_r}\}}^{[\star]}
\,  \prod_{s=1}^{n-m} \qop_{\{b_{k_{m+s}}\}}^{[n-m+1-2s]}
& m<n \\[10pt]
\epsilon^{k_1,...,k_m}
\prod_{r=1}^m \qop_{\{a_{k_r}, b_r\}}
& m=n \\[10pt]	%\det Q_{a|j}
    (-1)^{(n-m)n}
\epsilon^{k_1,...,k_m}
\prod_{r=1}^{n} 
\frac{1}{\Delta_{a_{k_r} b_r} }
\qop_{\{a_{k_r}, b_r\}}^{[\star]}
\,  \prod_{s=1}^{m-n} \qop_{\{a_{k_{n+s}}\}}^{[m-n+1-2s]} 
& m>n
\end{matrix} \right. \,,\no
\label{eq:detformulas}
\ee
see e.g. \cite{Tsuboi:2009ud,Tsuboi:2011iz,Gromov:2014caa}.
Here $|a_j|=0$ and $|b_j|=1$, $\qop^{[n]}=\qop(z+\sfrac{n}{2})$, and $\star$ can take any value in $-|m-n|,-|m-n|+2,...,|m-n|-2,|m-n|$. 
The prefactor is a consequence of the normalisation we use for the 
Q-operators \eqref{eq:qop}, cf. Appendix~\ref{sec:norm}.

The procedure to construct all Q-operators in this way
is shown in Figure~\ref{fig:solveQQ}. 
As a consequence of this construction, one finds that the Q-operators
only develop poles at $z\in\mathbb{N}$ or $z\in\mathbb{N}-\frac{1}{2}$, 
depending on the number of indices. 

\section{The BMN vacuum of fully twisted \nfsym at leading order} 
\label{sec:vacuum}

In this section we want to show how the Q-operator construction and the methods
for their evaluation can be applied to the \nfsym spin chain at the one-loop level.
To make comparisons to other approaches easier, we also show how to convert
our expressions to the conventions commonly used in the literature on the
Quantum Spectral Curve, see in particular \cite{Kazakov:2015efa}, where the twisted case is discussed.
From our construction we obtain the Q-operators for the theory with a
full diagonal twist. This generalises the well-know $\gamma_i$ and $\beta$ deformation
\cite{Leigh:1995ep,Lunin:2005jy,Frolov:2005dj}
and includes twists of the space-time part of the symmetries, such that 
the field theory is non-commutative.%
\footnote{
    See \cite{Beisert:2005if} for a discussion of the subtleties which
    arise when trying to deduce the precise non-commutative field theory
    from the integrable spin chain description.
}
The results can be specialised to the $\gamma_i$ and $\beta$ deformed cases,
or to the untwisted theory by choosing the twist angles appropriately.
While this leads to divergent matrix elements in the Q-operators, 
their eigenstates and the conserved charges such as the Hamiltonian which can be obtained from the Q-operators as described in \cite{Frassek:2012mg} remain finite.

To specialise our construction to \nfsym at one-loop, 
we first restrict to the singleton representation of $\mathfrak{u}(2,2|4)$
by choosing a grading and applying particle-hole transformation as
\begin{equation}
    (\gr{a})_{a=1}^8=(0,0,1,1,1,1,0,0)
    \eqncom \qquad
    (\omega_a)_{a=1}^8
    =
    (+1,+1,-1,-1,-1,-1,-1,-1)
    \eqncom
    \label{eq:qsc_grading_ph}
\end{equation}
and requiring that the central charge vanishes, i.e. $\ccc=0$.
Comparing with \eqref{eq:realoscs}, this gives the representation of
the fields of \nfsym in terms of the oscillators 
$(\oab_1,\oab_2,\odb_1,\odb_2,\odb_3,\odb_4,\obb_1,\obb_2)$
typically used in the spin chain description of \nfsym at weak coupling
and first investigated in \cite{Gunaydin:1984fk}. 
With this choice, the representation has the scalar field $\zop$ as
the lowest-weight state:
\begin{equation}
    \ket{\zop}=\odb_1\odb_2\ket{0}=\ket{0,0,1,1,0,0,0,0}
    \eqndot
\end{equation}

To facilitate the application of our results to \nfsym and to make comparisons
with the literature easier, we note that our conventions can easily be transformed 
into those typically employed by literature on the Quantum Spectral Curve of \nfsym.
There, bosonic and fermionic indices are treated separately. 
To obtain Q-functions with the expected asymptotics, we call the Q-operators of the lowest level
\begin{equation}
        (\qop_a)_{a=1}^8
    =
    (
    \qop_{\es|1},
    \qop_{\es|2},
    \qop_{1|\es,}
    \qop_{2|\es,}
    \qop_{3|\es,}
    \qop_{4|\es,}
    \qop_{\es|3,}
    \qop_{\es|4}
    )\,.
    \label{eq:qsc_qs}
\end{equation}
We note that the eigenvalues of these operators correspond to the leading perturbative contribution to the functions that appear in the $\mathbf{P}\mu$ and
$\mathbf{Q}\omega$ systems of the Quantum Spectral Curve, which govern the monodromy properties of the Q-system of $\mathcal{N}=4$ SYM at any coupling \cite{Gromov:2013pga,Gromov:2014caa}.
To obtain the twist variables which were used in \cite{Kazakov:2015efa}, we set
\begin{equation}
    (e^{-i\phi_a})_{a=1}^8
    =(\tw_a)_{i=a}^8
    =
    (y_1,y_2,x_1,x_2,x_3,x_4,y_3,y_4)\,.
    \label{eq:qsc_twists}
\end{equation}
Finally, the spectral parameter used in the QSC is related to the one used here by $z+\frac{1}{2} = i u$,
and the Lerch transcendents are given in terms of so-called $\eta$ functions, which in the twisted
case are defined by
\(\eta_a^x (u) \coloneqq \sum_{k=0}^\infty \frac{x^k}{(u+ik)^a}\). For the generalised Lerch
transcendents see Appendix~\ref{sec:lerch}.
These conventions ensure that the Q-operators have poles at positions in
the spectral parameter plane which are expected from the Quantum Spectral Curve.

As a simple application of the formulas derived in this paper, and in order 
to give some further examples of how they can be used in practise,
we calculate 
the matrix elements of the single-index Q-operators
with the BMN vacuum $\tr\zop^L$ of arbitrary length $L$.
Since these states constitute their own ``magnon blocks'',
we directly obtain the corresponding Q-functions in this case.
We consider the matrix elements of the single-index $\qlax$-operators
of the form $\bra\zop \qlax \ket\zop$. These can be determined from the integral representation 
of the diagonal part of the $\qlax$-operators given in 
\eqref{eq:middlepart_lowest};
equivalently one can use the finite sum representation in \eqref{eq:finite_sum_nonpolynomial} or \eqref{eq:middlepartlowest}.
Further relevant formulas are given
in Appendix~\ref{sec:matrixelements} where we describe the combinatorial
structure arising from the oscillator algebra, see in particular~\eqref{eq:melowest}.
For the matrix elements under consideration, there are in fact no 
combinatorial factors and no signs.
Since we look at matrix elements on the diagonal,
there are no auxiliary space operators,
which means that $m_A=\mt_A=\mmz_A$ in equation \eqref{eq:melowest}.
Thus we only have to evaluate the diagonal part given in \eqref{eq:middlepart_lowest},
where all $n_{\dt a}$ are zero.

We can now evaluate the integrals appearing in the diagonal part \eqref{eq:middlepart_lowest}.
The matrix elements of the operators $\qlax_{\{1\}},\ldots,\qlax_{\{6\}}$ 
are polynomials in the spectral parameter and in the number operators
in the auxiliary space, since all sums truncate. 
In this case the integral in \eqref{eq:middlepart_lowest}
is a contour integral which computes a residue, and can be evaluated by using
the series representations of hypergeometric functions ${}_2F_1$. 
The operators $\qlax_{\{7\}}$ and $\qlax_{\{8\}}$ have non-truncating sums
and their matrix elements are rational functions of both the spectral parameter
as well as the auxiliary space operators. For them, the integral in \eqref{eq:middlepart_lowest}
has to be taken along the interval $(0,1)$; using 
$\pFq{2}{1}{a,-b}{a}{x}=(1-x)^b$ and $\pFq{2}{1}{0,b}{c}{x}=1$, 
one directly finds Beta integrals, which give these rational functions.
Performing these calculations one finds the following matrix elements:
\begin{equation}
    \begin{aligned}
    \bra\zop\qlax_{\{1\}}\ket\zop &= 
    \bra\zop\qlax_{\{2\}}\ket\zop =
    \bra\zop\qlax_{\{3\}}\ket\zop =
    \bra\zop\qlax_{\{4\}}\ket\zop =1\,,
    \\[3pt]
    \bra\zop\qlax_{\{5\}}\ket\zop &= z+\frac{1}{2}+\NN_{51}+\NN_{52}+\NN_{53}+\NN_{54}\,,
    \\[3pt]
    \bra\zop\qlax_{\{6\}}\ket\zop &= z+\frac{1}{2}+\NN_{61}+\NN_{62}+\NN_{63}+\NN_{64}\,,
    \\[3pt]
    \bra\zop\qlax_{\{7\}}\ket\zop &= 
    \frac{1}{z+\frac{1}{2}-\NN_{71}-\NN_{72}-\NN_{73}-\NN_{74}}\,,
    \\[3pt]
    \bra\zop\qlax_{\{8\}}\ket\zop &= 
    \frac{1}{z+\frac{1}{2}-\NN_{81}-\NN_{82}-\NN_{83}-\NN_{84}}\,.
    \end{aligned}
    \label{eq:rvacuum}
\end{equation}

We now calculate the actual Q-functions as 
\begin{equation}
    \bra{\zop^L} \qop_{\{a\}}(z) \ket{\zop^L}
    =
    \tw_a^{-(-1)^{\gr{a}}z}
    \widehat\str 
    \left(
    \bra\zop
    \qlax_{\{a\}}(z)
    \ket\zop^L\right)
    \eqncom
\end{equation}
where the BMN vacuum state of length $L$ is $\ket{\zop^L}=\ket{\zop}^{\otimes L}$.
All formulas that are needed to evaluate the supertraces over the 
auxiliary Fock space are collected in Appendix~\ref{sec:str},
and can directly be applied to the matrix elements under consideration.
From 
\eqref{eq:rvacuum} we immediately see that
\begin{equation}
    \bra{\zop^L} \qop_{\{a\}}(z) \ket{\zop^L} =
    \tw_a^{-(-1)^{\gr{a}}z}
    \eqncom
    \qquad
    a=1,2,3,4
    \eqndot
\end{equation}
Using the multinomial theorem and the formula for the supertrace of polynomials
in the number operators given in \eqref{eq:str_polynomial}
we find
\begin{equation}
    \begin{aligned}
        &\bra{\zop^L} \qop_{\{5\}}(z) \ket{\zop^L}  = 
        \\&
        \tw_5^{z}
       \sum_{k=0}^L 
    z^k
    \Bigg[
        \sum_{\substack{k_0+k_1+k_2\\+k_3+k_4=L-k}}
        \binom{L}{k, k_0, k_1, k_2, k_3, k_4}
    \frac{
        \sum_{\ell_3=0}^{k_3} \eulerian{k_3}{\ell_3}\big(\frac{\tw_5}{\tw_3}\big)^{\ell_3+1-\delta_{k_3,0}}
        \sum_{\ell_4=0}^{k_4} \eulerian{k_4}{\ell_4}\big(\frac{\tw_5}{\tw_4}\big)^{\ell_4+1-\delta_{k_4,0}}
    }{
        2^{k_0} 
        \big(\frac{\tw_5}{\tw_5-\tw_1}\big)^{\delta_{k_1,0}-1}
        \big(\frac{\tw_2}{\tw_5-\tw_2}\big)^{\delta_{k_2,0}-1}
        \big(1-\frac{\tw_5}{\tw_3}\big)^{k_3}
        \big(1-\frac{\tw_5}{\tw_4}\big)^{k_4}
    }
    \Bigg]
    \eqncom
    \end{aligned}
\end{equation}
where we abbreviate the twist angles as $\tau_a=e^{-i\phi_a}$.
The Q-function $\bra{\zop^L} \qop_{\{6\}}(z) \ket{\zop^L} = \bra{\zop^L} \qop_{\{5\}}(z) \ket{\zop^L} \vert_{\tw_5\to \tw_6}$
is obtained by a simple relabelling of the result for $\qop_{\{5\}}$.
For the non-rational Q-functions 
we can use \eqref{eq:str_rational} to first evaluate the fermionic traces;
the first bosonic trace generates Lerch transcendents according to \eqref{eq:str_rational}.
The last trace can then be evaluated using \eqref{eq:str_hl}, and 
the resulting expressions simplified via the identity \eqref{eq:hl_shift}.
For $\qop_{\{7\}}$ we find
\begin{equation}
    \begin{aligned}
    \bra{\zop^L} \qop_{\{7\}}(z) \ket{\zop^L} = 
    \tw_7^{-z}
    \frac{(\tw_2-\tw_7)(\tw_1-\tw_7)}{(\tw_4-\tw_7)(\tw_3-\tw_7)}
    \Bigg[
    \frac{1}{(z+\frac{1}{2})^L}
    &
    +(-1)^L
        \frac{(\tw_2-\tw_3)(\tw_2-\tw_4)}{(\tw_1-\tw_2)\tw_2}
        {\textstyle \HL^{\tw_7/\tw_2}_L(-z-\frac{1}{2})}
       \\& 
       +(-1)^L
        \frac{(\tw_1-\tw_3)(\tw_1-\tw_4)}{(\tw_2-\tw_1)\tw_1}
        {\textstyle \HL^{\tw_7/\tw_1}_L(-z-\frac{1}{2})}
    \Bigg]\,.
    \end{aligned}
\end{equation}
The calculation proceeds similarly for $\qop_{\{8\}}$,
and gives $\bra{\zop^L} \qop_{\{8\}}(z) \ket{\zop^L} = \bra{\zop^L} \qop_{\{7\}}(z) \ket{\zop^L} \vert_{\tw_7\to \tw_8}$.
Quite remarkably, the Q-functions for these most trivial states of the theory
are already rather complicated, due to the presence of the full twist.
The higher level Q-functions for the BMN vacuum can be generated from the 
ones given above as described in Section~\ref{sec:abcde}, using
\eqref{eq:finite_diff} and \eqref{eq:detformulas}.
We note that the calculations for excited states are not much more difficult;
the corresponding Q-operators for each magnon block can likewise be evaluated
using the formulas in Appendix~\ref{sec:formulas}.

\section{Conclusions and Outlook}
\label{sec:conclusion}

In this article, we discussed the oscillator construction of the Baxter Q-operators
of integrable models for the case of non-compact super spin chains with representations
of Jordan-Schwinger form, focusing on the concrete evaluation of these operators.
We outlined the derivation of the Lax operators on which this construction is based,
and defined the Q-operators with their functional relations.
For non-compact spin chains with infinite-dimensional state spaces,
these Lax operators are given in terms of
infinite sums which hides the analytic properties of the resulting Q-system and complicates their evaluation.
We proposed a strategy to overcome these difficulties.
For the Lax operators of the lowest level, we derived a representation
without infinite sums, which allows to compute explicit matrix elements.
Employing a small set of formulas for the normalised supertrace, 
it is then possible to obtain the matrix elements of the corresponding
Q-operators. Due to the remaining symmetry, the Q-operators can be realised
as finite matrices for each magnon block, and the functional relations
then allow to uniquely recover the entire Q-system starting from the lowest level.
For all the steps in this procedure we provided explicit formulas
which can directly be implemented in computer algebra systems for practical 
calculations.

Although our approach only relies on the Q-operators of the lowest level to determine 
the whole Q-system, it would be desirable to find analogues of our
integral formula \eqref{eq:middlepart_lowest} also for the Lax operators of higher levels.
Our initial studies in Section~\ref{sec:higher} indicate that this rather
difficult task might require novel ideas.
A promising route might be to derive these formulas directly from the 
Yang-Baxter equation \eqref{eq:yberll}.
Our approach naturally incorporates compact spin chains with symmetric representations at each site of the quantum space.
It would furthermore be interesting to study whether it can be generalised to more general representations and in particular to principal series representations. Furthermore, it should be straightforward to apply our method to open spin chains, for which the study
of Baxter Q-operators was initiated only recently in \cite{Frassek:2015mra}.
The main motivation of our work is to allow the application of Q-operators to concrete physical problems. 
Apart from applications in high energy physics, we hope that our method can be applied in the context of the ODE/IM correspondence \cite{Dorey:2007zx} and the computation of correlation functions~\cite{Boos:2006mq,Boos:2008rh,Jimbo:2008kn}.

    Currently our main focus 
lies on \nfsym where a similar Q-system arises in the form of the Quantum Spectral Curve. 
So far, the QSC of \nfsym has only been investigated on the eigenvalue level, and it is tempting to ask how it lifts to the operatorial level, see also the discussion in \cite{Staudacher:2010jz}. The individual Q-functions are multivalued functions of the spectral parameter with particular monodromies and asymptotic behaviour. These Q-functions are believed to be eigenvalues of Q-operators, but the nature of the operatorial Q-system remains a mystery. Our approach should be equivalent to the construction of the leading perturbative contribution to this system. It is well-understood how to iteratively construct perturbative corrections to the Q-functions \cite{Marboe:2014gma,Gromov:2015vua}. There is no immediate reason why these methods should not lift to the operatorial level, even though the perturbative solution of the QSC with general twists, where eigenstates possibly correspond to spin chain states with non-zero momentum, has not yet been examined in detail in the literature. 
% Our initial studies show that the presence of six independent twists very quickly make these calculations extremely complicated. Given 
Thus a systematic way of performing the untwisting would be of great practical
value. On the eigenvalue level, a rather general method was proposed in
\cite{Kazakov:2015efa}, but it has to be applied to each state individually; a discussion
on the operator level can be found in \cite{Korff2006,Bazhanov:2010ts} for the case of the Heisenberg
spin chain, see also \cite{Pronko:1998xa}. 
%Furthermore, solutions of the QSC corresponding to non-zero spin chain momentum have a more involved structure, which has not yet been fully treated in the literature. 
% For general twists all solutions are of this type, but one can potentially overcome this by imposing extra constraints on the twists and restricting to a subset of the Hilbert space.
Such computations would yield access to perturbative information about the operatorial Q-system, which might give hints about its deeper nature. Furthermore, diagonalisation of the Q-operators would immediately yield higher loop corrections to the eigenstates of the dilatation operator. This information might shed light on higher-point correlation functions about which many aspects are still not fully understood and also on the emergence of the integrable system that underlies \nfsym from the field theory. We plan to address these questions in future work.

It would furthermore be interesting to apply our construction 
to \nfsym in the presence of defects \cite{deLeeuw:2015hxa,Buhl-Mortensen:2015gfd}
where Q-operators have already been used to calculate one-point functions,
and to the integrable chiral field theories discovered recently \cite{Gurdogan:2015csr}.

We finally note that there are other approaches to the construction
of Q-operators, which superficially are quite distinct from 
the oscillator construction pursued in this work.
It would be interesting to see how results analogous to those presented here can be obtained from the approach developed in \cite{Derkachov:2006fw,Derkachov:2010qe},
and if the interesting Q-operator construction in \cite{Kazakov:2010iu} can be generalised to non-compact
spin chains.

\section*{Acknowledgements}
We like to thank Matthias Staudacher, Dmytro Volin, Zengo Tsuboi, Gregory Korchemsky, Gregor Richter, Leonard Zippelius, Ivan Kostov, Didina Serban, 
and Stijn van Tongeren for interesting discussions. RF thanks Vasily Pestun for related discussions.
We thank the referees for useful remarks.
Further we thank the IPhT, Saclay and \emph{``Mathematische Physik von Raum, Zeit und Materie''}, Humboldt University Berlin for hospitality.
DM received support from GK 1504 \emph{``Masse, Spektrum, Symmetrie''}.
RF is supported by the IH\'{E}S visitor program.
The research leading to these results has received funding from the People Programme
(Marie Curie Actions) of the European Union’s Seventh Framework Programme FP7/2007-
2013/ under REA Grant Agreement No 317089 (GATIS).

\appendix

\section{Formulas for the explicit evaluation of Q-systems} 
\label{sec:formulas}

\subsection{Matrix elements of lowest level $\qlax$-operators}
\label{sec:matrixelements}

Explicit matrix elements
$\bra{\mtb} \qlax_{\{a\}}(z) \ket{\mb}$
of the $\qlax$-operators of the lowest level
with the states defined in \eqref{eq:defstates}
can be obtained using simple oscillator algebra
from \eqref{eq:form_lax_lowest} with the 
diagonal part $\middlepart_{\{a\}}$ given by the
integral representation \eqref{eq:middlepart_lowest}, 
or equivalently by the finite sum formulas \eqref{eq:finite_sum_nonpolynomial}.
First note that the values of the summation variables
$n_{\dt a}$, $\dt a \in \bar I$ in \eqref{eq:form_lax_lowest} 
are fixed by the difference in occupation numbers
\begin{equation}
    n_{\dt a} = -\omega_{\dt a}(\mt_{\dt a}-m_{\dt a}) \,,
    \label{eq:fixns_one}
\end{equation}
because each of the corresponding oscillators only appears in a single factor 
$X_{\udt a \dt a}$ and a single factor $Y_{\udt a \dt a}$ in \eqref{eq:form_lax_lowest}.
The powers of the oscillators with index $\udt a$ in the left and right factors are
\begin{equation}
    N_{\ell} = \sum_{\dt a} \theta(n_{\dt a}) \abs{n_{\dt a}}
    \qquad \text{and} \qquad
    N_{r} = \sum_{\dt a} \theta(-n_{\dt a}) \abs{n_{\dt a}}
    \eqndot
\end{equation}
Then the occupation numbers of the 
state on which the diagonal part $\middlepart_{\{a\}}$ acts are given by
\begin{equation}
    \mmz_{\udt a} = \mt_{\udt a}-\omega_{\udt a} N_\ell 
    = m_{\udt a}-\omega_{\udt a}N_r
    \eqncom
    \qquad
    \mmz_{\dt a} =
    \begin{cases}
        \max(m_{\dt a},\mt_{\dt a})  & \text{if } \omega_{\dt a} = 1 \\
        \min(m_{\dt a},\mt_{\dt a}) & \text{if } \omega_{\dt a} = -1 
    \end{cases}
    \eqndot
    \label{eq:m0s_one}
\end{equation}
For the different types of oscillators, given in \eqref{eq:realoscs}, this sets
the number operators in the diagonal part $\middlepart_{\{a\}}$ to
\begin{equation}
    \NN_c=\left\{\begin{array}{rrrl}
        \oab_c\oa_c &\rightarrow &\mmz_c &\qquad \gr{c}=0\text{ and }\omega_c=+1 \\
        -1-\obb_c\ob_c  &\rightarrow &-1-\mmz_c &\qquad \gr{c}=0\text{ and }\omega_c=-1 \\
        \ocb_c\oc_c &\rightarrow& \mmz_c &\qquad \gr{c}=1\text{ and }\omega_c=+1 \\
        1-\odb_c\od_c &\rightarrow& 1-\mmz_c &\qquad \gr{c}=1\text{ and }\omega_c=-1 \\
    \end{array}\right.
    \eqndot
\end{equation}
Finally, collecting all combinatorial factors arising from the oscillator algebra,
we can write the matrix elements as
\begin{equation}
\begin{aligned}
    &\bra{\mtb}
    \qlax_{\{\udt a\}}(z) 
    \ket{\mb}
    =\\[6pt]&\;\;\;
    \oxb_{\udt a 1}^{\theta(-\omega_{1}(\mt_{1}-m_{1}))\abs{-\omega_{1}(\mt_{1}-m_{1})}}
    \cdots
    \oxb_{\udt a K}^{\theta(-\omega_{K}(\mt_{K}-m_{K}))\abs{-\omega_{K}(\mt_{K}-m_{K})}}
    \\[6pt]&\;\;\;
    (-1)^{\sum_{\dt a} \abs{n_{\dt a}} c_{\udt a \dt a}}
    \left( 
        \frac{\mmz_{\udt a} !}{\sqrt{\mt_{\udt a}!m_{\udt a}!}} 
    \right)^{-\omega_{\udt a}}
    \prod_{\dt a}\sqrt{\frac{\max(m_{\dt a},\mt_{\dt a}) !}{\min(m_{\dt a},\mt_{\dt a})!}}
    \;\;
     \middlepart_{\{a\}}(z, \{\mmz\}, \{ -\omega_{\dt a}(\mt_{\dt a}-m_{\dt a}) \}) 
     \\[6pt]&\;\;\;
    \ox_{K \udt a}^{\theta(\omega_{K}(\mt_{K}-m_{K}))\abs{-\omega_{K}(\mt_{K}-m_{K})}}
    \cdots
    \ox_{1 \udt a}^{\theta(\omega_{1}(\mt_{1}-m_{1}))\abs{-\omega_{1}(\mt_{1}-m_{1})}}\,,
\end{aligned}
\label{eq:melowest}
\end{equation}
where $K=p+q+r+s$ for $\mathfrak{u}(p,q|r+s)$ models and the sign is determined by
\begin{equation}
    \footnotesize
    \begin{aligned}
        c_{\udt a \dt a}&=
        \Big[ 
            (\gr{\udt a}+\gr{\udt a}\gr{\dt a})\theta(n_{\dt a})+(1+\gr{\udt a}\gr{\dt a})\theta(-n_{\dt a})
        \Big]
        +
        \sfrac{1}{2}\Big[ 
            (\gr{\udt a}+1)(1-\omega_{\udt a})
            \theta(n_{\dt a})
            +
            (\gr{\dt a}+1)(1-\omega_{\dt a})
            \theta(-n_{\dt a})
        \Big]
        \\
        &+
        \Big[ 
            (\gr{\udt a}+\gr{\dt a})
            \sum_{c=1}^{K} \gr{c} 
            \big(
                \tilde{m}_c \theta(n_{\dt a})
                +{m}_c \theta(-n_{\dt a})
            \big)
        \Big]
        +
        \Big[\gr{\udt a}
            \sum_{1\leq c<\udt a}  \gr{c} 
            \big(
                \tilde{m}_c\theta(n_{\dt a})
                +{m}_c\theta(-n_{\dt a})
                +\delta_{c\dt a}
            \big)
        \Big]
        \\
        &+
        \Big[
            \gr{\dt a}\sum_{1\leq c<\dt a}  \gr{c} 
            \big(
                (\theta(n_c)+\tilde{m}_c)\theta(n_{\dt a})
                +(\theta(-n_c)+{m}_c)\theta(-n_{\dt a})
            \big)
        \Big]\,,
    \end{aligned}
    \label{eq:uglysigns}
\end{equation}
where we set $n_{\udt a}=0$.

\subsection{Calculating supertraces}
\label{sec:str}

    For the Q-operators of the 
    lowest level, the following formulas are sufficient to perform
    the occurring sums.
    First, for polynomials in the number operators, we can use
    \begin{equation}
        \widehat\str_{ab}\; \NN_{ab}^k = \begin{cases}
            \frac{
                \sum_{n=0}^k \eulerian{k}{n} 
                \big( \frac{\tw_a}{\tw_b} \big)^{n+1-\delta_{k,0}}
            }{
                \big( 1-\frac{\tw_a}{\tw_b} \big)^k
            } & \text{bosonic}
            \\[20pt]
            \Big( \frac{\tw_a}{\tw_a-\tw_b} \Big)^{1-\delta_{k,0}} & \text{fermionic}
        \end{cases}\;,
        \label{eq:str_polynomial}
    \end{equation}
    where $\eulerian{k}{n}$ are the Eulerian numbers defined by
    \begin{equation}
        \eulerian{k}{n}
        =
        \sum_{j=0}^{n+1} (-1)^j \binom{k+1}{j} (n-j+1)^k
        \eqndot
        \label{eq:eulerian}
    \end{equation}
    Here and in the following we abbreviate the twist angles via $\tw_a=\exp(-i\phi_a)$.

    For the non-rational Q-operators, we also need the Lerch transcendent $\HL$
    defined in \eqref{eq:def_hl}, since the matrix elements of the
    $\qlax$-operators are rational functions of the number operators.
    Concretely one encounters traces of the form
   \begin{equation}
        \widehat\str_{ab} \;
        \frac{\NN_{ab}^k}{(\NN_{ab}+r)^\ell} = 
        \begin{cases}
            \frac{\tw_b-\tw_a}{\tw_b}
            \sum_{m=0}^k\binom{k}{m}(-r)^{k-m}\HL^{\tw_a/\tw_b}_{\ell-m}(r)
            &\qquad\text{bosonic}
            \\[10pt]
            \frac{1}{\tw_b-\tw_a}
            \Big(
            \delta_{k,0} \tw_b \frac{1}{r^\ell} - \tw_a \frac{1}{(r+1)^\ell}
            \Big)
            &\qquad\text{fermionic}
        \end{cases}
        \eqndot
        \label{eq:str_rational}
    \end{equation}
    If further traces have to be evaluated, summation formulas for the 
    Lerch transcendent are
    \begin{equation}
        \widehat\str_{ab} \; \HL^\tw_\ell(\NN_{ab}+r)  = \begin{cases}
            \frac{\tw_b-\tw_a}{\tw \tw_b -\tw_a}
            \Big(
            \tw \HL^\tw_\ell(r) - \frac{\tw_a}{\tw_b}\HL^{\tw_a/\tw_b}_\ell(r)
            \Big)
            &\qquad\text{bosonic}
            \\[10pt]
            \frac{\tw_b}{\tw_b-\tw_a}
            \Big(
               \tw_b \HL^\tw_\ell(r) - \tw_a \HL^\tw_\ell(r+1)
            \Big)
            &\qquad\text{fermionic}
        \end{cases} \;,
        \label{eq:str_hl}
    \end{equation}
    and in the general case
    \begin{equation}
    \begin{aligned}
        &\widehat\str_{ab} \; \NN_{ab}^k\HL^\tw_\ell(\NN_{ab}+r)
        \\[15pt]
        &= 
        \begin{cases}
            \frac{\tw_b-\tw_a}{\tw_b}\Bigg\{
    \frac{
        \delta_{k,0}+\sum_{t=1}^k \eulerian{k}{t-1}\big(\frac{\tw_a/\tw_b}{\tw}\big)^t
    }{
        \big(1-\frac{\tw_a/\tw_b}{\tw}\big)^{k+1}
    }
    \HL^\tw_\ell(r)
    -
    \frac{1}{\tw}
    \sum_{s=0}^k \binom{k}{s}
    \frac{
        \delta_{s,k}+\sum_{t=1}^{k-s}\eulerian{k-s}{t-1}\big(\frac{\tw_a/\tw_b}{\tw}\big)^{t}
        }{
            \big( 1- \frac{\tw_a/\tw_b}{\tw}\big)^{k-s+1}
        }
        &
        \\ \qquad\times
        \bigg[
            \bigg(
            \sum_{j=0}^s\binom{s}{j}(1-r)^{s-j}\HL^{\tw_a/\tw_b}_{\ell-j}(r-1)
            \bigg)
            -\delta_{s,0}\frac{1}{(r-1)^\ell}
        \bigg]
    \Bigg\}
           &\!\!\!\!\!\!   \!\!\!\!\!\! \!\!\!\!\!\! \!\!\!\!\!\! \!\!\!\!\!\!\text{bosonic}
           \\[20pt]
           \frac{1}{\tw_b-\tw_a}
           \Big(
           \tw_b \delta_{k,0}\HL^\tw_\ell(r)
           - \tw_a \HL^\tw_\ell(r+1)
           \Big)
           &\!\!\!\!\!\! \!\!\!\!\!\! \!\!\!\!\!\! \!\!\!\!\!\! \!\!\!\!\!\! \text{fermionic}
       \end{cases} \,.
    \end{aligned}
    \end{equation}

    It is evident that these types of supertraces are the only ones which
    can appear. This implies that all matrix elements of the lowest level
    Q-operators are either rational functions for truncating $\qlax$-operators or can be written in terms of the Lerch transcendent for non-truncating $\qlax$-operators.
    We note that one can usually reduce the number of terms containing this function using the identity
    \begin{equation}
        \HL^\tw_k(z+1) = \frac{1}{\tw} \bigg(
        \HL^\tw_k(z)-\frac{1}{z^k} \bigg)
        \eqndot
        \label{eq:hl_shift}
    \end{equation}

\subsection{Generalised Lerch transcendents}
\label{sec:lerch}

When solving difference equations in order to obtain the Q-operators with one
bosonic and one fermionic index, cf.~Section~\ref{sec:abcde}, we need to apply the discrete integration $\Sigma$.
It can be realised as $\Sigma[f(z)]=\sum_{k=0}^\infty f(z+k)$ when the sum is convergent.
For non-rational Q-operators, these sums lead to generalisations of
the Lerch transcendent \eqref{eq:def_hl}.
The treatment here is equivalent to that of $\eta$-functions given in \cite{Marboe:2014gma,Gromov:2015dfa}.

We define
\begin{equation}
\HL_{a_1,a_2,...,a_n}^{\tw_1,\tw_2,...,\tw_n}(z)=\sum_{0\le k_1<k_2<...<k_n}^\infty \frac{\tw_1^{k_1}\tw_2^{k_2}\cdots \tw_n^{k_n}}{(z+k_1)^{a_1}(z+k_2)^{a_2}\cdots(z+k_n)^{a_n}}
\eqncom
\end{equation} 
where the number of parameters $n$ is arbitrary. Note that while the variables $\tau_i$
will be given in terms of the twist variables in concrete calculations, 
here their indices $i$ are not $\mathfrak{gl}(N|M)$ indices, but simply label these
arbitrary arguments.

It is clear from the definition that the generalised Lerch transcendent 
satisfies the following shift identity, generalising equation \eqref{eq:hl_shift}:
\be
\HL_{a,a_1,\ldots,a_n}^{\tw,\tw_1,\ldots,\tw_n}(z)
=\frac{\tw_1\cdots\tw_n}{z^a}\HL_{a_1,\ldots,a_n}^{\tw_1,\ldots,\tw_n}(z+1)
+\tw\tw_1\cdots\tw_n \HL_{a,a_1,\ldots,a_n}^{\tw,\tw_1,\ldots,\tw_n}(z+1) \eqndot
\label{eq:HLshift} 
\ee
Importantly, $\HL$-functions satisfy so called stuffle-relations, e.g.
\be
\HL_{a_1}^{\tw_1} \HL_{a_2}^{\tw_2} 
= \HL_{a_1,a_2}^{\tw_1,\tw_2} 
+ \HL_{a_1+a_2}^{\tw_1\tw_2}
+\HL_{a_2,a_1}^{\tw_2,\tw_1}
\eqndot
\ee
These can be used to linearise all products of these functions. 
The Lerch transcendents
are related to $\eta$-functions used in the Quantum Spectral Curve literature by
\be
\HL_{a_1,a_2,...,a_n}^{\tw_1,\tw_2,...,\tw_n}(z)
= i^{a_1+a_2+\hdots+a_n}%{\sum_{j=1}^n a_j}
\eta_{a_1,a_2,...,a_n}^{\tw_1,\tw_2,...,\tw_n}(iz)
\eqndot
\ee

\subsection{Formulas for discrete integrals}
\label{sec:Psi}
To generate the full Q-system,
we need to apply the discrete integration $\Sigma$ to the following four classes of functions,
which form a closed set.

\paragraph{Polynomials}
In the case $\tw\neq 1$, $\Sigma(\tw^z z^a)$ is another polynomial 
with an overall exponential factor of the form 
$p(z)=\tw^z(c_a z^a+...+c_0)$ satisfying the constraint $p(z)-p(z+1)=\tw^z z^a$.
This constraint fixes $p(z)$ completely. In the %untwisted 
case %with 
$\tw=1$, 
the polynomial is of degree $a+1$ instead.

\paragraph{Shifted inverse powers}
From the definition of the generalised Lerch transcendent one finds
\be
\Sigma\left[ \frac{\tw^z}{(z+m)^a} \right] = \tw^z \HL_a^\tw(z+m)
\eqndot
\ee

\paragraph{Terms of the form $\frac{\tw^z\HL}{(z+m)^a}$}
Note that
\be
\Sigma\left[\frac{
    \tw^z \HL_{a_1,a_2,...,a_n}^{\tw_1,\tw_2,...,\tw_n}(z+1)
  }{
    z^a
}\right]
  =\tw^z
  \HL_{a,a_1,a_2,...,a_n}^{\tw,\tw_1,\tw_2,...,\tw_n}(z) 
  \eqndot
\label{eq:HLneg}
\ee
To evaluate $\Sigma\left[\frac{\tw^z\HL_{a_1,a_2,...,a_n}^{\tw_1,\tw_2,...,\tw_n}
}{(z+m)^a}\right]$, use \eqref{eq:HLshift} to align the shifts and then use \eqref{eq:HLneg}.

\paragraph{Terms of the form $\tw^z z^a \HL$}
To evaluate products of monomials and generalised Lerch transcendents,
one can use  the finite difference analogue of partial integration,
\be
\Sigma\left[ f \left(g-g^{[2]}\right) \right] = f g -\Sigma\left[ g^{[2]} \left(f-f^{[2]}\right) \right]
\eqndot
\ee
For $\Sigma[\tw^z z^a  \HL_{a_1,a_2,...,a_n}^{\tw_1,\tw_2,...,\tw_n}]$, 
set $f=(\tw_1\cdots\tw_n)^z  \HL_{a_1,a_2,...,a_n}^{\tw_1,\tw_2,...,\tw_n}$ and 
$g-g^{[2]}=\left(\frac{\tw}{\tw_1\cdots\tw_n}\right)^z z^a$. 
This can be used recursively until no terms of this type are present.

\section{Highest level $\qlax$-operators and their matrix elements}
\label{sec:highest}

In Section~\ref{sec:lowest} we derived a representation of $\qlax$-operators
for the lowest level of the Q-system which allows to evaluate their matrix
elements. 
Here we summarise similar results for the $\qlax$-operators of the highest 
non-trivial level, i.e. those where the index set $I$ contains all but one single
index which we denote by $\dt a$.

According to equation \eqref{eq:multireorder}
we can write these $\qlax$-operators as
\begin{equation}
    \qlax_{\overline{\{\dt a\}}}(z) =
    \sum_{\{n_{\udt a}\}=-\infty}^\infty
    \left[
        \prod_{\udt a} 
        (Y_{\udt a \dt a})^{\theta(+n_{\udt a})\abs{n_{\udt a}}}
    \right]
    \middlepart_{\overline{\{\dt a\}}}(z,\{\NN\}, \{ n\})
    \left[
        \prod_{\udt a} 
        (-X_{\udt a \dt a})^{\theta(-n_{\udt a})\abs{n_{\udt a}}}
    \right]
    \eqncom
    \label{eq:form_lax_highest}
\end{equation}
where $X$ and $Y$ are given in \eqref{eq:defxy}. After performing
an almost identical computation as for the $\qlax$-operators with a single index,
the diagonal part can be written as
\begin{equation}
    \begin{aligned}
    &\middlepart_{\overline{\{\dt a\}}}(z,\{\NN\}, \{ n\})
    =
    -
    \frac{\Gamma( z+1+\frac{1}{2}(-1)^{\gr{\dt a}}   )}{
        \Gamma( z+1-c-\frac{1}{2}(-1)^{\gr{\dt a}}  )}
    \\&\times
    \int \dd t \;
    (-t)^{\NN_{\dt a}+(-1)^{\gr{\dt a}}-1}
    (1-t)^{-z-1-\frac{1}{2}(-1)^{\gr{\dt a}}}
    \prod_{\udt a}
    \frac{1}{\abs{n_{\udt a}}!}
    \pFq{2}{1}{-\NN_{\udt a}, -\NN_{\udt a \dt a}}{1+\abs{n_{\udt a}}}{(-1)^{\gr{\dt a}+\gr{\udt a}}t}
    \eqndot
    \end{aligned}
    \label{eq:middlepart_highest}
\end{equation}
Here the integration again depends on whether the oscillator with index $\dt a$
is particle-hole transformed or not,
\begin{equation}
    \int \dd t = \begin{cases}\displaystyle
        \frac{1}{\Gamma(\NN_{\dt a}+(-1)^{\gr{\dt a}})} \int_0^1 \dd t 
        & \quad \text{if } \gr{\udt a}=0 \text{ and } \omega_{\udt a} = +1 \\[20pt]
        \displaystyle
        (-1)^{\NN_{\dt a}+1} \frac{\Gamma(1-\NN_{\dt a}-(-1)^{\gr{\dt a}})}{2\pi i} \oint_{t=0} \dd t
        & \quad \text{else}
    \end{cases}
    \eqndot
    \label{eq:def_int_lowest}
\end{equation}

We can easily obtain matrix elements 
\(
    \bra{\mtb}
    \qlax_{\overline{\{\dt a\}}}(z) 
    \ket{\mb}
\)
from this representation.
First we note that occupation numbers $\mtb$ and $\mb$ fix the values of the
summation variables $n_{\udt a}$, $\udt a\in I$ in \eqref{eq:form_lax_highest} to
\(
    n_{\udt a} = \omega_{\udt a}(\mt_{\udt a}-m_{\udt a})
\).
The powers of the $\dt a$ oscillators in the left and right factors are
   $ N_{\ell} = \sum_{\udt a} \theta(n_{\udt a}) \abs{n_{\udt a}} $ and
   $ N_{r} = \sum_{\udt a} \theta(-n_{\udt a}) \abs{n_{\udt a}} $,
such that the occupation numbers of the states on which the diagonal part acts
are 
\begin{equation}
    \mmz_{\dt a} = \mt_{\dt a}+\omega_{\dt a} N_\ell 
    = m_{\dt a}+\omega_{\dt a}N_r
    \qquad
    \mmz_{\udt a} =
    \begin{cases}
        \min(m_{\udt a},\mt_{\udt a}) & \text{if } \omega_{\udt a} = 1 \\
        \max(m_{\udt a},\mt_{\udt a}) & \text{if } \omega_{\udt a} = -1 
    \end{cases}
    \eqndot
    \label{eq:m0s}
\end{equation}
We can thus write the matrix elements as
\begin{equation}
\begin{aligned}
    &\bra{\mtb}
    \qlax_{\overline{\{\dt a\}}}(z) 
    \ket{\mb}
    =
\\[6pt]
&\;\;
    \oxb_{1\dt a}^{\theta(\omega_{1}(\mt_{1}-m_{1}))\abs{\omega_{1}(\mt_{1}-m_{1})}}
    \cdots
    \oxb_{K\dt a}^{\theta(\omega_{K}(\mt_{K}-m_{K}))\abs{\omega_{K}(\mt_{K}-m_{K})}}
    \\[6pt]
&\;\;
    (-1)^{\sum_{\udt a} \abs{n_{\udt a}} c'_{\udt a \dt a}}
    \left( 
        \frac{\mmz_{\dt a} !}{\sqrt{\mt_{\dt a}!m_{\dt a}!}} 
    \right)^{\omega_{\dt a}}
    \prod_{\udt a}\sqrt{\frac{\max(m_{\udt a},\mt_{\udt a})  !}{ \min(m_{\udt a},\mt_{\udt a}) !}}
    \;\;
    \middlepart(z, \{\mmz\}, \{ \omega_{\udt a}(\mt_{\udt a}-m_{\udt a})
\}) 
\\[6pt]
&\;\;
    \ox_{\dt a K}^{\theta(-\omega_{K}(\mt_{K}-m_{K}))\abs{\omega_{K}(\mt_{K}-m_{K})}}
    \cdots
    \ox_{\dt a 1}^{\theta(-\omega_{1}(\mt_{1}-m_{1})\abs{\omega_{1}(\mt_{1}-m_{1})}}
\end{aligned}\,,
\end{equation}
where $K=p+q+r+s$ for $\mathfrak{u}(p,q|r+s)$ models and the sign follows from
\begin{equation}
    \footnotesize
    \begin{aligned}
        c'_{\udt a \dt a}&=
        % explicit signs in X and Y after commuting the oscillators a bit:
        \Big[ 
            (\gr{\udt a}+\gr{\udt a}\gr{\dt a})\theta(n_{\udt a})+(1+\gr{\udt a}\gr{\dt a})\theta(-n_{\udt a})
        \Big]
        %\\&
        +
        % from particle hole
        \sfrac{1}{2}\Big[ 
            (\gr{\udt a}+1)(1-\omega_{\udt a})
            \theta(n_{\udt a})
            +
            (\gr{\dt a}+1)(1-\omega_{\dt a})
            \theta(-n_{\udt a})
        \Big]
        \\
        &+
        % hopping with xis over states
        \Big[ 
            (\gr{\udt a}+\gr{\dt a})
            \sum_{c=1}^{K} \gr{c} 
            \big(
                \tilde{m}_c \theta(n_{\udt a})
                +{m}_c \theta(-n_{\udt a})
            \big)
        \Big]
        %\\&
        +
        % acting on the states
        \Big[ \gr{\dt a}
            \sum_{1\leq c<\dt a} \gr{c} 
            \big(
                \tilde{m}_c\theta(n_{\udt a})
                +{m}_c\theta(-n_{\udt a})
                +\delta_{c\udt a}
            \big)
        \Big]
        \\
        &+
        \Big[
           \gr{\udt a}   \sum_{1\leq c<\udt a}\gr{c} 
            \big(
                (\theta(n_c)+\tilde{m}_c)\theta(n_{\udt a})
                +(\theta(-n_c)+{m}_c)\theta(-n_{\udt a})
            \big)
        \Big]
        \eqndot
    \end{aligned}
    \label{eq:uglysigns2}
\end{equation}

\section{Normalisations of Q-operators and functional relations}
\label{sec:norm}

There are different conventions for the functional relations of Q-operators and twisted
Q-functions which have been used in the literature. These are related to
the normalisation of the Q-operators. 
To facilitate comparisons to other works, we summarise and compare these conventions in
this appendix. The twists will be parametrised by $\tw_a=e^{-i\phi_a}$.

In this work, we use the normalisation which is typically employed in
the literature on the oscillator construction of Q-operators, see e.g. \cite{Frassek:2010ga,Frassek:2011aa}.
The operators are defined in \eqref{eq:qop}, and the prefactor there can 
be written as
$\prod_{\udt a\in I}\tw_{\udt a}^{-(-1)^{\gr{\udt a}}z}$.
Note that this normalisation is compatible with indexing the operators by sets
(which is natural for the oscillator construction), since it does not impose
an ordering of the $\mathfrak{gl}(N|M)$ indices.
The functional relations are given in \eqref{eq:QQb} and \eqref{eq:QQf},
and involve the factors $\Delta_{ab}= (-1)^{\gr{a}} \frac{\tw_b-\tw_a}{\sqrt{\tw_a\tw_b}}$.

Other possible choices induce an ordering of the
$\mathfrak{gl}(N|M)$ indices. This ordering can be reflected by indexing
the Q-operators with antisymmetric multi-indices; we will instead
label the Q-operators with sets, and keep track of the ordering
in the functional relations.
One possibility, discussed on the level of Q-functions for example in \cite{Kazakov:2015efa}, uses a normalisation
without exponential scaling factors:
\begin{equation}
    \hat\qop_I(z)= 
    \prod_{\substack{\udt a, \udt b \in I \\ \udt a < \udt b}}
    (\tw_{\udt a}-\tw_{\udt b})^{(-1)^{\gr{\udt a}+\gr{\udt b}}}
    \,
    \widehat\str\,  \mathcal{M}_I(z)
    \eqncom
\end{equation} 
which gives functional relations
\begin{equation}
    \begin{aligned}
    &\hat\qop_{I\cup \{a,b\}} \hat\qop_I = 
    \tw\hat\qop_{I\cup \{a\}}^+ \hat\qop_{I\cup \{b\}}^-
    -\tilde\tw\hat\qop_{I\cup\{a\}}^- \hat\qop_{I\cup \{b\}}^+
    & \quad|a|=|b| \\
    &\hat\qop_{I\cup \{a\}} \hat\qop_{I\cup \{b\}} = 
    \tw\hat\qop_{I\cup\{a,b\}}^+ \hat\qop_{I}^-
    -\tilde\tw\hat\qop_{I\cup \{a,b\}}^- \hat\qop_{I}^-
    & \quad|a|\neq|b|
    \end{aligned}\;,
\end{equation}
where $\tw=\tw_a$ and $\tilde\tw=\tw_b$ if $a<b$
or  $\tw=\tw_b$ and $\tilde\tw=\tw_a$ if $b<a$,
and we used the notation
$f^{\pm}=f(z\pm\frac{1}{2})$.

Another normalisation we want to discuss is given by
\begin{equation}
    \check\qop_I(z)= 
    \prod_{\udt a\in I}\tw_{\udt a}^{(-1)^{\gr{\udt a}}(z+s_I)}  
    \prod_{\substack{\udt a, \udt b \in I \\ \udt a < \udt b}}
    (\tw_{\udt a}-\tw_{\udt b})^{(-1)^{\gr{\udt a}+\gr{\udt b}}}
    \,\widehat\str\,  \mathcal{M}_I(z)\,.
\end{equation} 
Here the shift $s_I$ is the one defined in \eqref{eq:def_shift}.
The functional relations for these operators are identical to those
of untwisted Q-functions:
\begin{equation}
    \begin{aligned}
    &\check\qop_{I\cup \{a,b\}} \check\qop_I = 
    \check\qop_{I\cup \{a\}}^+ \check\qop_{I\cup \{b\}}^-
    -\check\qop_{I\cup\{a\}}^- \check\qop_{I\cup \{b\}}^+
    & \quad|a|=|b| \\
    &\check\qop_{I\cup \{a\}} \check\qop_{I\cup \{b\}} = 
    \check\qop_{I\cup\{a,b\}}^+ \check\qop_{I}^-
    -\check\qop_{I\cup \{a,b\}}^- \check\qop_{I}^-
    & \quad|a|\neq|b|
    \end{aligned}\;.
\end{equation}
In the first equation, we have to assume that $a<b$ if $|a|=|b|=0$,
or $b<a$ if $|a|=|b|=1$; otherwise, the left hand side changes its sign. 

{
\small
\bibliographystyle{utphys2}
\bibliography{refs}
}

\end{document}